\documentclass{aastex631}

\usepackage{tabularx}
\usepackage{newtxtext,newtxmath}
\usepackage[T1]{fontenc}
\usepackage{eucal}
\usepackage{nicefrac}
\usepackage{amssymb}
\usepackage{latexsym}
\usepackage{amsmath, mathrsfs, bm}
\usepackage{eqnarray}
\usepackage{url}
\usepackage{cases}
\usepackage{epsfig}
\usepackage{color}
\usepackage{graphicx}
\usepackage{grffile}
\usepackage{float}
\usepackage{soul}
\usepackage{enumerate}
\usepackage{multirow}
\usepackage{threeparttable}
\usepackage{xspace}
\usepackage{bm}
\usepackage{times}
\usepackage{etoolbox}
\usepackage{multirow}

\usepackage{textcomp}

\usepackage{hyperref}
\begin{document}

\title{Comprehensive Statistical Analysis of Initial Lorentz Factor and Jet Opening Angle of Gamma-Ray Bursts}

\author[0009-0008-6979-2559]{Jian Zhang}
\affiliation{College of Physics and Electronic Information Engineering, Guilin University of Technology, Guilin 541004, China}

\author[0009-0009-2321-8305]{Bao-Cheng Qin}
\affiliation{College of Physics and Electronic Information Engineering, Guilin University of Technology, Guilin 541004, China}

\author[0000-0003-0726-7579]{Lu-Lu Zhang}
\affiliation{College of Physics and Electronic Information Engineering, Guilin University of Technology, Guilin 541004, China}

\author[0000-0002-5936-8921]{Fu-Wen Zhang}
\affiliation{College of Physics and Electronic Information Engineering, Guilin University of Technology, Guilin 541004, China}
\affiliation{Key Laboratory of Low-dimensional Structural Physics and Application, Education Department of Guangxi Zhuang Autonomous Region, Guilin 541004, China}

\correspondingauthor{Fu-Wen Zhang}
\email{fwzhang@pmo.ac.cn}

\begin{abstract}
The initial Lorentz factor ($\Gamma_{0}$) and jet half-opening angle ($\theta_{\rm jet}$) of gamma-ray bursts (GRBs) are critical physical parameters for understanding the dynamical evolution of relativistic jets and the true energy release of GRBs. We compile a sample of 89 GRBs that exhibit an onset bump feature in their early optical or GeV light curves, 42 of which also display a jet break feature, and derive their $\Gamma_{0}$ and $\theta_{\rm jet}$ values. Using this sample, we re-eaxmine the correlations between $\Gamma_{0}$ and the prompt emission parameters (isotropic energy $E_{\rm iso}$, peak luminosity $L_{\rm iso}$, and peak energy $E_{\rm p}$). Our results confirm the previously reported $\Gamma_{0}$$-$$E_{\rm iso}$ ($L_{\rm iso}$), $\Gamma_{0}$$-$$E_{\rm p,z}$, and $E_{\rm iso}$ ($L_{\rm iso}$)$-$$E_{\rm p}$$-$$\Gamma_{0}$ relations for both homogeneous interstellar medium (ISM) and wind density profiles (Wind). Notably, we find that the short GRB 090510 complies with the $\Gamma_{0}$$-$$E_{\rm iso}$ relation, but significantly deviates from the $\Gamma_{0}$$-$$L_{\rm iso}$ and $\Gamma_{0}$$-$$E_{\rm p,z}$ relations. We systematically investigate the influence of $\theta_{\rm jet}$ on $\Gamma_{0}$ and find a weak dependence, which reads $\Gamma_{0}$ (ISM) $\propto$ $\theta^{-0.18 \pm 0.15}_{\rm jet}$ (ISM) and $\Gamma_{0}$ (Wind) $\propto$ $\theta^{-0.72 \pm 0.14}_{\rm jet}$ (Wind). Additionally, we report, for the first time, three new three-parameter correlations, i.e., $E_{\rm iso}$ $\propto$ $\theta^{-0.52 \pm 0.26}_{\rm jet}$ (ISM) $\Gamma^{2.21 \pm 0.28}_{0}$ (ISM) and $E_{\rm iso}$ $\propto$ $\theta^{-0.65 \pm 0.21}_{\rm jet} $ (\text{Wind}) $\Gamma^{2.08 \pm 0.19}_{0}$ (Wind); $L_{\rm iso}$ $\propto$ $\theta^{-0.35 \pm 0.27}_{\rm jet}$ (ISM) $\Gamma^{2.46 \pm 0.29}_{0}$ (ISM) and $L_{\rm iso}$ $\propto$ $\theta^{-0.29 \pm 0.28}_{\rm jet} $(\text{Wind}) $\Gamma^{2.37 \pm 0.25}_{0}$ (Wind); $E_{\rm p,z}$ $\propto$ $\theta^{-0.15 \pm 0.19}_{\rm jet}$ (ISM) $\Gamma^{1.11 \pm 0.21}_{0}$ (ISM) and $E_{\rm p,z}$ $\propto$ $\theta^{-0.17 \pm 0.23}_{\rm jet} $ (\text{Wind}) $\Gamma^{1.09 \pm 0.21}_{0}$ (Wind). These tight $E_{\rm iso}$ (or $L_{\rm iso}$, $E_{\rm p,z}$)$-$$\Gamma_{0}$$-$$\theta_{\rm jet}$ correlations likely arise from the combined effects of radiation mechanisms, jet structure, and outflow dynamics in GRBs. In addition, we further explore the relations between the initial Lorentz factor and the jet-corrected energy, and find $\Gamma_{0}$ (ISM) $\propto$ $E^{0.22 \pm 0.04}_{\gamma,52}$ (ISM) and $\Gamma_{0}$ (Wind) $\propto$ $E^{0.40 \pm 0.06}_{\gamma,52}$ (Wind); $\Gamma_{0}$ (ISM) $\propto$ $L^{0.20 \pm 0.04}_{\gamma,52}$ (ISM) and $\Gamma_{0}$ (Wind) $\propto$ $L^{0.30 \pm 0.06}_{\gamma,52}$ (Wind). We also find that the jet-corrected correlations remain significant, suggesting that these relations are intrinsic to the physical nature of GRBs. However, the increased dispersion after correction implies that underlying differences persist among individual GRBs.

\end{abstract}


\section{Introduction} \label{sec:intro}
Gamma-ray bursts (GRBs) are among the most energetic explosions in the Universe \citep{1993ApJ...413..281B,1993ApJ...413L.101K,2002ApJ...581.1236Z}.
Since their discovery in the 1960s, numerous models have been proposed to understand their underlying mechanisms, among which the fireball shock model is the most widely accepted.
In this framework, the GRB fireball evolves through three stages: acceleration, coasting, and deceleration \citep{1999ApJ...513..669K}.
During the acceleration phase, the fireball is primarily driven by radiation pressure, and the Lorentz factor $\Gamma$ increases with the accelerated expansion of the fireball, i.e., $\Gamma$ $\propto$ $r$, where $r$ is the radius of the fireball.
As the fireball transitions into the coasting phase, the Lorentz factor saturates at its maximum value, referred to as the initial Lorentz factor ($\Gamma_{0}$), and remains nearly constant. 
Only when the fireball interacts with the circumburst medium and enters the deceleration phase does $\Gamma_{0}$ begin to decline significantly.
The non-thermal energy spectra observed in GRBs provide compelling evidence that their ejecta propagate at ultra-relativistic speeds. 
Direct confirmation of this relativistic motion came from \citet{2004ApJ...609L...1T}, who tracked the radio afterglow of GRB 030329 and measured its bulk velocity to be in the range of 3$c$$-$5$c$. 
The evolution of the Lorentz factor directly reflects the relativistic motion of the fireball, and systematic investigation of the statistical properties of $\Gamma_0$ is crucial for a deeper understanding the mechanism of GRBs.

Since $\Gamma_{0}$ cannot be directly measured, several estimation methods have been proposed.
The most commonly used method is based on the deceleration time of the relativistic blastwave (the onset time of the afterglow) \citep{2018pgrb.book.....Z}, during which a smooth onset bump is expected to appear in the early afterglow light curve as a signature of the fireball’s deceleration. In this scenario, $\Gamma_0$ can be directly derived from the peak time ($t_{\rm p}$) of the bump \citep{1993ApJ...405..278M,2003ApJ...595..950Z,2007A&A...469L..13M}.
The second method is based on the compactness argument, where the high-energy cutoff in the prompt emission spectrum is interpreted as a signature of $e^{+}e^{-}$ pair production, thereby constraining the jet's Lorentz factor.
This method requires the observation of high-energy photons (e.g., GeV photons) \citep{1997ApJ...491..663B,2001ApJ...555..540L,2008MNRAS.384L..11G}.
The third method relies on the very early external shock radiation, which forms during the prompt emission phase but remains undetected due to its low luminosity relative to the prompt emission.
As a result, only a lower limit on the Lorentz factor can be derived \citep{2010MNRAS.402.1854Z}.

With the growing sample of measured $\Gamma_0$, the intrinsic characteristics between $\Gamma_0$ and prompt emission have been extensively studied.
Initially, \citet{2010ApJ...725.2209L} studied 20 GRBs with deceleration features in their early optical afterglows and found a tight correlation between $\Gamma_{0}$ and the isotropic energy $E_{\rm iso}$, i.e., $\Gamma_0 \propto E_{\rm iso,52}^{0.25 \pm 0.03}$.
\citet{2012ApJ...751...49L} confirmed the $\Gamma_0$$-$$E_{\rm iso}$ relation using a sample of 38 GRBs and also found a similar relation between $\Gamma_0$ and the average luminosity $L_{\rm iso}$ ($L_{\rm iso}$ = (1+z) $E_{\rm iso}$/$T_{\rm 90}$), i.e., $\Gamma_0$ $\propto$ $L^{0.30 \pm 0.002}_{\rm iso,52}$.   
Subsequently, \citet{2015ApJ...813..116L} introduced the peak energy $E_{\rm p}$ as a third key parameter and obtained a tight three-parameter correlation among $L_{\rm iso}$, $E_{\rm p,z}$ and $\Gamma_{0}$, i.e., $L_{\rm iso}$ $\propto$ $E^{1.34 \pm 0.14}_{\rm p,z}$$\Gamma^{1.32 \pm 0.19}_{0}$, where $E_{\rm p,z} = E_{\rm p} (1+z)$.  
Meanwhile, \citet{2019ApJ...876...77X} proposed using the $L_{\rm iso}$-$E_{\rm p,z}$-$\Gamma_0$ relation to estimate the pseudo-$\Gamma_0$ of Long GRBs (LGRBs) and studied LGRBs properties in the comoving frame.
Furthermore, \citet{2012MNRAS.420..483G,2018A&A...609A.112G} further analyzed the $\Gamma_{0}$$-$$E_{\rm iso}$ ($L_{\rm iso}$) and $\Gamma_{0}$$-$$E_{\rm p,z}$ relations for both homogeneous interstellar medium (ISM) and wind density profiles (Wind).

It was generally assumed that GRB radiation is isotropic.
However, with the launch of the Swift satellite, its rapid response and precise positioning capabilities revealed achromatic temporal breaks in the afterglow light curves of some GRBs.
This suggests that the outflows of GRBs should be collimated rather than isotropic \citep{1997ApJ...487L...1R}.
In the fireball external shock scenario, the burst ejecta moves with a relativistic speed
and is assumed to form a conical jet with the half-opening angle $\theta_{\rm jet}$.
As the ejecta interacts with circumburst medium and decelerates, the relativistic beaming angle 1/$\Gamma$ will increase with time \citep{2018ApJ...859..160W,2023ApJ...950...30Z}.
When the condition 1/$\Gamma$ $>$ $\theta_{\rm jet}$ is satisfied, a steepening break is predicted to appear in the afterglow light curve (known as the jet break), with the corresponding time is called jet break time $T_{\rm jet}$ \citep{2001Natur.414..853W,2018pgrb.book.....Z}.
Current observational data indicate that this time break feature is not only present in the optical band but has also been observed in the X-ray and/or radio band.

The deployment of Swift/XRT and ground-based telescopes has significantly expanded the multi-wavelength GRB afterglow database.
Building on this foundation, numerous statistical studies focusing on jet breaks have been conducted.
For instance, \citet{2001ApJ...562L..55F} and \citet{2003ApJ...594..674B} corrected the isotropic energy and found that the geometrically corrected energy $E_{\gamma}$ ($E_{\rm \gamma} = E_{\rm iso}(1-\cos\theta_{\rm jet})$) follows a very narrow distribution, primarily  clustered around 5 $\times 10^{50}$$-$10$^{51}$ erg.
\citet{2004ApJ...616..331G} expanded the sample of GRBs with jet break features and found a tight correlation between $E_{\rm p,z}$ and $E_{\rm \gamma}$, i.e., $E_{\rm p,z}$ $\propto$ $E^{0.7}_{\gamma}$.
\citet{2005ApJ...633..611L} utilized a sample of 15 GRBs with jet breaks in the optical band and found a tight three-parameter $E_{\rm iso}$$-$$E_{\rm p,z}$$-$$T_{\rm j,z}$ relation, i.e., $E_{\rm iso,52}$ $\propto$ $E^{1.94 \pm 0.17}_{\rm p,z}$$T^{-1.24 \pm 0.23}_{\rm j,z}$, where $T_{\rm j,z} = T_{\rm jet}/(1+z)$, which was then applied to constrain cosmological parameters. 
More recently, \citet{2020ApJ...900..112Z} further confirmed the $E_{\rm iso}$$-$$E_{\rm p,z}$$-$$T_{\rm j,z}$ relation using a sample of 138 GRBs with jet breaks in X-ray, optical, and radio afterglows, and discovered a new $E_{\rm jet}$$-$$T_{\rm j,z}$$-$$E_{\rm p,z}$ relation, i.e., $E_{\rm jet}$ (ISM) $\propto$ $T^{0.67 \pm 0.02}_{\rm j,z}$$E^{0.84 \pm 0.07}_{\rm p,z}$ and $E_{\rm jet}$ (Wind) $\propto$ $T^{0.45 \pm 0.01}_{\rm j,z}$$E^{0.56 \pm 0.05}_{\rm p,z}$.

With the continuous expansion of the known $\Gamma_{0}$ sample, we aim to investigate whether $\Gamma_{0}$ follows the empirical relations derived from previous studies. Especially for GRBs samples where both $\Gamma_{0}$ and $\theta_{\rm jet}$ are available, this enables a systematic exploration of the intrinsic relations involved in GRBs prompt emission and their underlying physical nature.
The structure of this paper is organized as follows. 
The sample selection and data processing are described in Section 2.
The statistical results are presents in Section 3.
Our conclusions are summarized in Section 4.
Throughout this paper, the cosmological parameters $H_{\rm 0}$ = 71 km s$^{-1}$ Mpc$^{-1}$, $\Omega_{\rm M}$ = 0.27, and $\Omega_{\rm \Lambda}$ = 0.73 are adopted. 
We also use the notation $Q_n = {Q}/{10^n}$.

\section{Sample Selection and Data Analysis} \label{sec:data}
To systematically investigate the relations between $\Gamma_{0}$, $\theta_{\rm jet}$, and prompt emission parameters (isotropic energy $E_{\rm iso}$, peak luminosity $L_{\rm iso}$, and rest frame peak energy $E_{\rm p,z}$) of GRBs, we collect 89 GRBs (including 88 long GRBs and 1 short GRB) that exhibit a distinct onset bump in their early optical afterglow or GeV light curves, where 86 GRBs have well measured redshifts. The majority of the $t_{\rm p}$ values in the sample are adopted from \citet{2010ApJ...725.2209L, 2013ApJ...774...13L} and \citet{2018A&A...609A.112G} (see Table \ref{tab:tab1}).
Among them, the $t_{\rm p}$ values of 8 GRBs are derived from GeV light curves and are marked with an “L” in Table \ref{tab:tab1}.

For 3 bursts without measured redshifts, we adopt $z = 2$, which corresponds to the median value of the measured redshift distribution (the short GRB 090510 is excluded).
Additionally, we identify 42 GRBs exhibiting jet break features from the sample, characterized by a normal decay segment (slope $\sim$ $-$1) followed by a steeper decay phase (slope $\sim$ $-$2).  
The corresponding jet break times were mainly collected from \citet{2012ApJ...745..168L}, \citet{2020ApJ...900..112Z}, and the Swift-XRT webpage \footnote{\url{https://www.swift.ac.uk/xrt_live_cat/}} (see Table \ref{tab:tab2} for details).

Among the 89 GRBs, 53 are detected by Fermi/GBM or Konus/Wind instruments. Their fluence, flux, and spectral parameters (including the spectral indexes $\alpha$ and $\beta$, and the peak energy $E_{\rm p}$) are obtained from the Fermi online catalog\footnote{\url{https://heasarc.gsfc.nasa.gov/W3Browse/fermi/fermigbrst.html}} and the Konus-Wind online catalog\footnote{\url{https://www.ioffe.ru/LEA/catalogs.html}}. 
The remaining 36 GRBs are detected by Swift/BAT. Of these, only one burst has a spectrum well fitted by a cutoff power-law (CPL) model. 
For the other 35 GRBs, a simple power-law (PL) model is applied, and thus their $E_{\rm p}$ values cannot be directly obtained.
For 17 events, the $E_{\rm p}$ values are adopted from \citet{2015ApJ...813..116L} and \citet{2025ApJ...Sun}.
For the remaining 18 GRBs, $E_{\rm p}$ is estimated using the empirical relation proposed by \citet{2007ApJ...655L..25Z}. 

Based on the spectral parameters obtained above, we calculated the isotropic energy $E_{\rm iso}$ and peak luminosity $L_{\rm iso}$ of all GRBs. 
To minimize the influence of different detectors on the results, we apply the following procedure to the spectral indices: (1) For the CPL model, the value of $\alpha$ is retained, while $\beta$ is fixed at the typical Band function value of -2.25 \citep{2000ApJS..126...19P}.
(2) For the PL model, both $\alpha$ and $\beta$ are fixed at the typical Band function values of $\alpha = -1.0$ and $\beta = -2.25$ \citep{2000ApJS..126...19P}.
Furthermore, both $E_{\rm iso}$ and $L_{\rm iso}$ are corrected to the 1$-$10$^{4}$ keV energy band in the rest frame.
The $E_{\rm iso}$ is estimated by
\begin{equation}  
E_{\rm iso} = \frac{4 {\pi} D^{2}_{\rm L} S_{\rm \gamma}}{(1 + z)}k , \label{eq:equation1} 
\end{equation} 
where $S_{\rm \gamma}$ is the fluence (in units of erg\,cm$^{-2}$), $D_{\rm L}$ is the luminosity distance, and $k$ is the $k$-correction factor.
The isotropic peak luminosity $L_{\rm iso}$ is estimated as
\begin{equation}  
L_{\rm iso} = 4 {\pi} D^{2}_{\rm L} F_{\rm p} k , \label{eq:equation2} 
\end{equation} 
where $F_{\rm p}$ is the peak flux (in units of erg\,cm$^{-2}$\,s$^{-1}$ or photons\,cm$^{-2}$\,s$^{-1}$).

We use the same formula derived by \citet{2014MNRAS.445.1625N} to calculate the initial Lorentz factor $\Gamma_{0}$ based on the onset bump peak time $t_{\rm p}$ of afterglow, which can be written as 

\begin{equation}  
\Gamma_0 = \left[\frac{(17-4s)}{16\pi(4-s)}\left(\frac{E_0}{n_{0} m_{\rm p} c^{(5-s)}}\right)\right]^{1/(8-2s)} t_{\rm p,z}^{-\frac{3-s}{8-2s}}, \label{eq:equation3}  
\end{equation} 
where $E_{0}$ is the isotropic equivalent kinetic energy, $n_{0}$ is the medium number density, $m_{\rm p}$ is the proton rest mass, and $t_{\rm p,z}$ is the peak time in the rest frame, $t_{\rm p,z} = {t_{\rm p}}/{(1+z)}$.

We derive the values of $\Gamma_{0}$ for both a homogeneous density ISM ($s = 0$) and a wind density profile ($s = 2$).
For ISM case, we adopt the typical value $n_{0} =$ 1 cm$^{-3}$.
For Wind case, $n_{0}$ = $\dot{M}$/{4$\pi r^{2}{m_{\rm p}}{v_w}$}, where $\dot{M}$ is the mass-loss rate and $v_{w}$ is the wind velocity, with typical values of $\dot{M}$ = $10^{-5}$$M_{\odot}$\,yr$^{-1}$ and $v_w$ = $10^{3}$ km\,s$^{-1}$, respectively.
In both cases, the isotropic equivalent kinetic energy can be estimated by $E_{0}$ = $E_{\rm iso}$/$\eta$, where a radiative efficiency $\eta$ = 20$\%$ is adopted.

The jet half-opening angle $\theta_{\rm jet}$ of GRBs can be estimated from the jet break time $T_{\rm jet}$, which is defined as
\begin{equation}
\begin{aligned}
\theta_{\rm jet}\,(\rm ISM) & = 0.076 \, \text{rad} \left( \frac{T_{\text{\rm jet}}}{1 \, \text{day}} \right)^{{3/8}} \left( \frac{1 + z}{2} \right)^{-{3/8}}  \\
& \times \, E_{\text{iso}, 53}^{-{1/8}} \left( \frac{\eta}{0.2} \right)^{{1/8}} \left( \frac{n}{1 \, \text{cm}^{-3}} \right)^{{1/8}}, 
\label{eq:equation4}
\end{aligned}
\end{equation}
for a homogeneous density ISM \citep{1999ApJ...525..737R,1999ApJ...519L..17S,2001ApJ...562L..55F,2015ApJ...807...92Y}, and
\begin{equation}
\begin{aligned}
\theta_{\rm jet}\,(\rm Wind) & = 0.12 \, \text{rad} \left( \frac{T_{\text{jet}}}{1 \,\text{day}} \right)^{{1/4}} \left( \frac{1 + z}{2} \right)^{-{1/4}} \\
& \times \, E_{\text{iso}, 52}^{-{1/4}} \left( \frac{\eta}{0.2} \right)^{{1/4}} A_{*}^{{1/4}},
\label{eq:equation5}
\end{aligned}
\end{equation}
for a wind density profile \citep{2000ApJ...536..195C,2003ApJ...594..674B,2015ApJ...807...92Y}.
For ISM case, we adopt the typical value $n =$ 1 cm$^{-3}$.
For Wind case, the wind parameter $A_{*}$ = 1 is adopted. 
Meanwhile, the radiative efficiency is adopted as $\eta$ = 20$\%$ in both cases.

All the errors of these quantities are calculated by the error
propagation formula.
The values of $z$, $E_{\rm iso}$, $L_{\rm iso}$, $E_{\rm p,z}$ and $\Gamma_{0}$ for each GRB are listed in Table \ref{tab:tab1}.
Table \ref{tab:tab2} presents the values of $T_{\rm jet}$ and $\theta_{\rm jet}$ for the GRBs exhibiting jet break features.

\section{Results} \label{ssec:Res}
\subsection{\texorpdfstring{Distributions of $\Gamma_{0}$ and $\theta_{\rm jet}$}{Distributions of Gamma0 and thetajet}} \label{subsec:dist}

\begin{figure*}
\centering
\includegraphics[angle=0,scale=0.50]{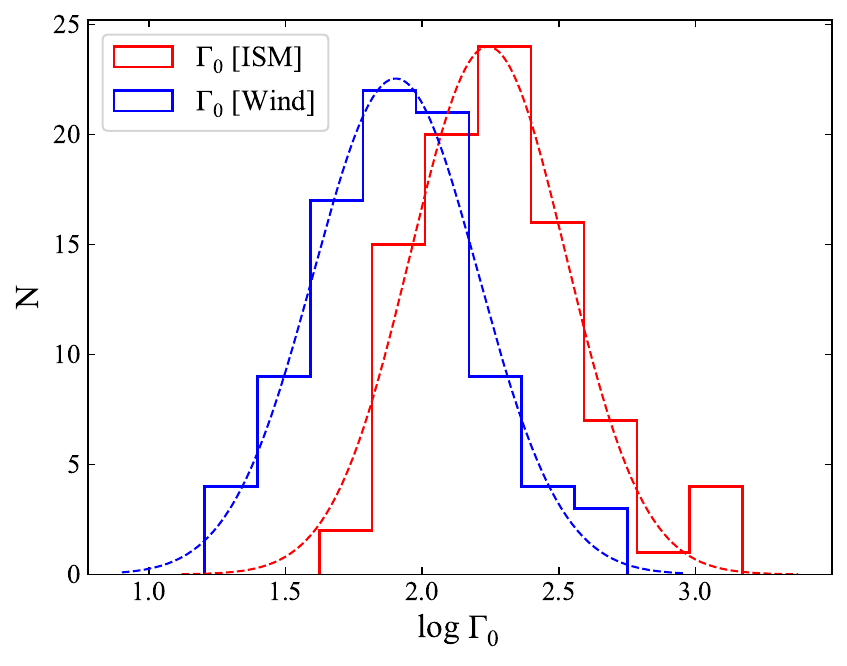}
\includegraphics[angle=0,scale=0.50]{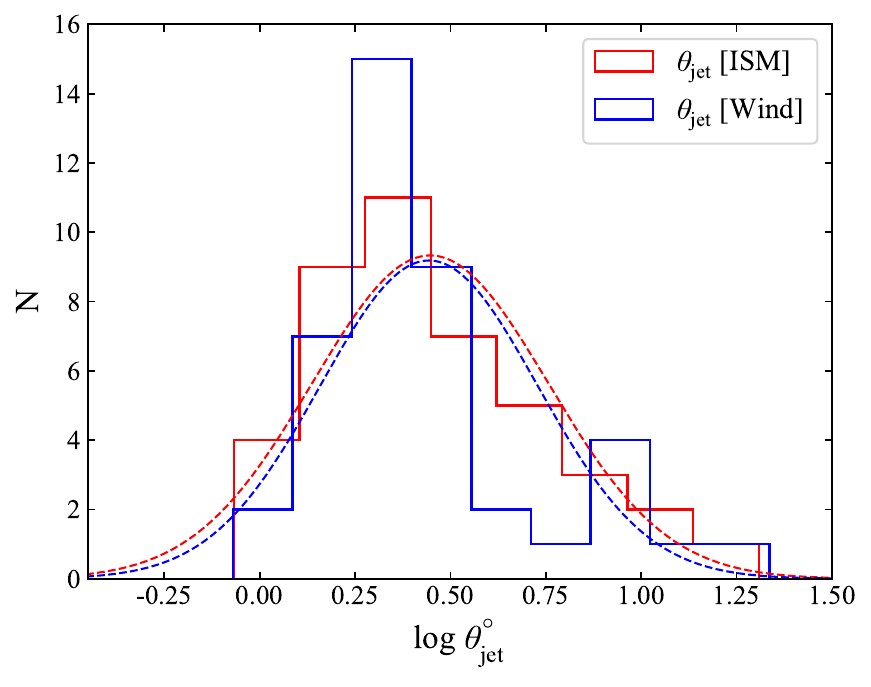}
\caption{Distributions of the initial Lorentz factor $\Gamma_{0}$ (Left panel) and the jet half-opening angle $\theta_{\rm jet}$ (Right panel). The red and blue lines represent the ISM and wind cases, respectively. The dotted lines indicate the best-fitting curves. 
\label{fig:figure1}}
\end{figure*}

We analyze the distributions of $\Gamma_{0}$ and $\theta_{\rm jet}$. 
The distribution of $\Gamma_{0}$ is shown in the left panel of Figure \ref{fig:figure1}, where the red and blue stepped lines represent the cases of a homogeneous ISM and a wind density profile, respectively.
The values of $\Gamma_{0}$ span a wide range from tens to thousands, ranging from 42 to 1486 in the ISM case and from 16 to 563 in the wind case, both approximately following a log-normal distribution.
To obtain a quantitative result, we perform a fit using a Gaussian function.
From the best-fit results, we obtain that the median values of $\Gamma_{0}$ for two different medium cases are $\Gamma_{0}$ $=$ $174^{{}+335}_{{}-90}$ (ISM) and $\Gamma_{0}$ $=$ $80^{{}+160}_{{}-40}$ (Wind). 
All the errors are 1$\sigma$ uncertainties. 
Obviously, the median $\Gamma_{0}$ in the homogeneous medium is approximately twice that in the wind medium.
Additionally, an interesting feature is that among the 8 GRBs with $t_{\rm p}$ obtained from GeV light curves, 4 events have $\Gamma_{0}$ values exceeding 1000 in the ISM case.
For the 81 GRBs with $t_{\rm p}$ derived from early optical afterglow light curves, 16 have $\Gamma_{0}$ below 100 in the ISM medium, and 55 fall below this threshold in the wind medium.
However, none of these 81 GRBs exhibit $\Gamma_{0}$ values above 1000 in either medium.

\citet{2018A&A...609A.112G} previously calculated the values of $\Gamma_{0}$ for the bursts with detected optical onset bumps for both ISM and wind medium cases, and obtained $\Gamma_{0} = 178^{{}+240}_{{}-142}$ for ISM case and  $\Gamma_{0}$ $= 85^{{}+117}_{{}-65}$ for wind case.
We find the results are consistent within the permissible systematic error range.
More recently, \citet{2024ApJ...972..170Z} performed a comprehensive analysis of 30 GRBs exhibiting textbook-version light curve.
Using a standard forward shock model to jointly fit the optical and X-ray light curves, they obtained a mean initial Lorentz factor of $\Gamma_{0} = 426^{{}+794}_{{}-229}$ for the ISM medium. The discrepancy is mostly due to the difference of $n_{0}$ and $\eta$ values.

The distribution of $\theta_{\rm jet}$ is displayed in the right panel of Figure \ref{fig:figure1}, with the homogeneous ISM and wind density profiles represented by red and blue stepped lines, respectively. 
As illustrated in the figure, the $\theta_{\rm jet}$ distributions exhibit remarkable consistency between the ISM (0.86\textdegree{}$-$20.33\textdegree{}) and wind (0.85\textdegree{}$-$21.66\textdegree{}) cases. 
Gaussian fitting yields median values of $2.80^{{}+5.70}_{{}-1.38}$\textdegree{} (ISM) and $2.78^{{}+5.35}_{{}-1.44}$\textdegree{} (Wind), respectively. 
\citet{2018ApJ...859..160W} revisited the jet breaks using a sample of 55 GRBs. Applying the closure relation, they estimated $\theta_{\rm jet}$ for each bursts, finding values spanning 0.36\textdegree{}$-$7.39\textdegree{} with a mean of 2.5\textdegree{}$\pm$1.0\textdegree{}.
It is evident that our estimated $\theta_{\rm jet}$ values in both the ISM and wind profiles span a broader range compared to the results of \citet{2018ApJ...859..160W}, while the mean values remain comparable.

\subsection{Multivariate Correlations of the Initial Lorentz Factor with Prompt Emission Parameters} \label{ssec:gamma_corre}

\begin{figure*}
\centering
\includegraphics[angle=0,scale=0.50]{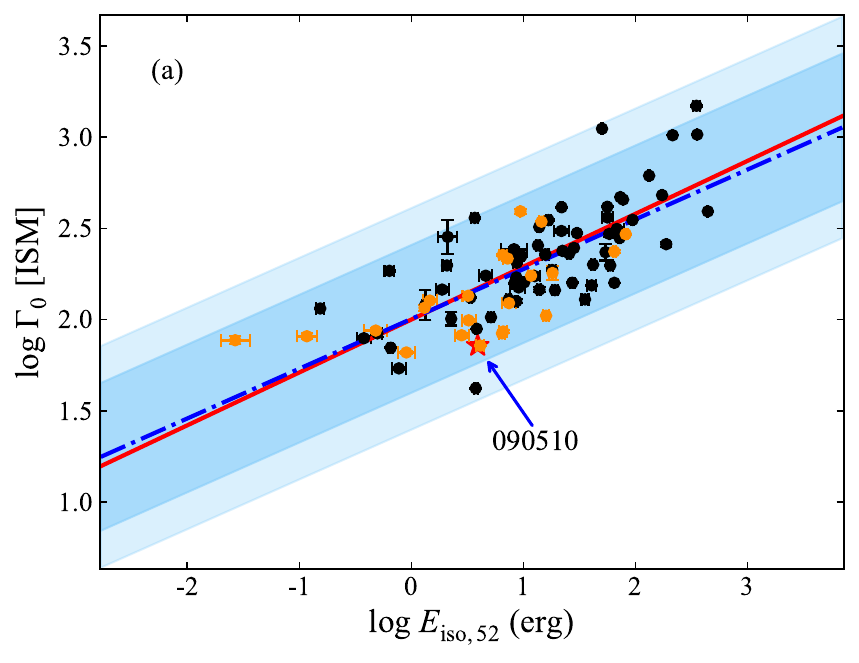}
\includegraphics[angle=0,scale=0.50]{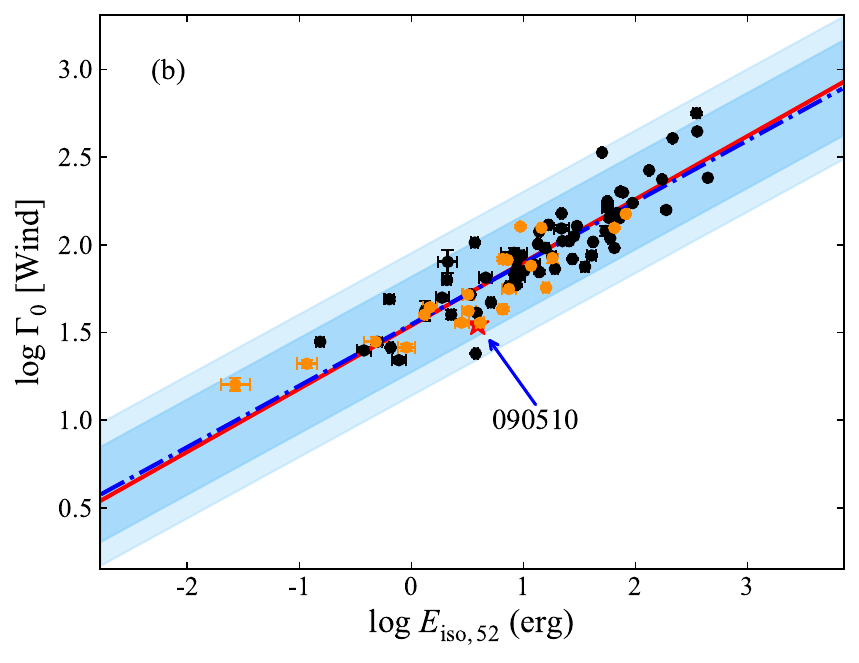}
\includegraphics[angle=0,scale=0.50]{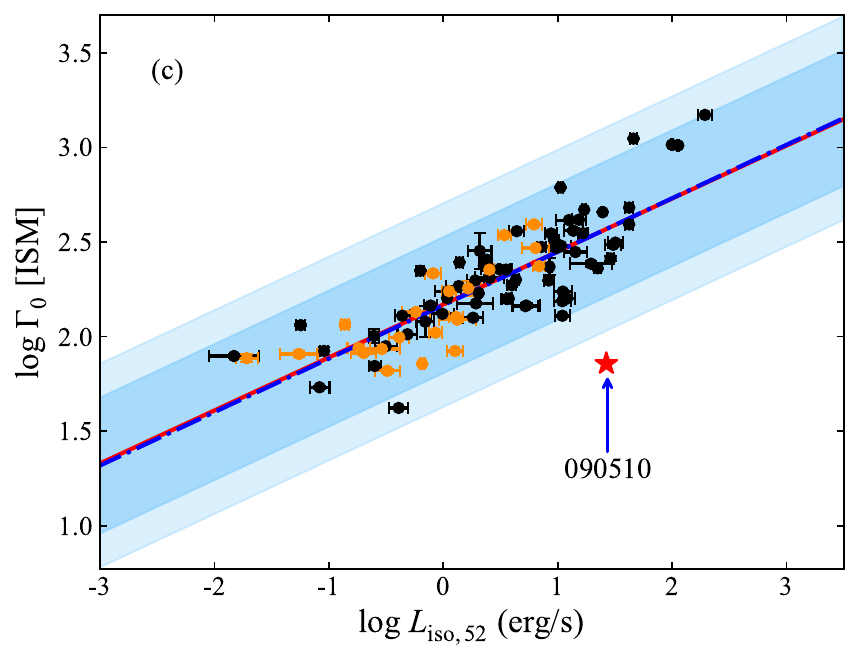}
\includegraphics[angle=0,scale=0.50]{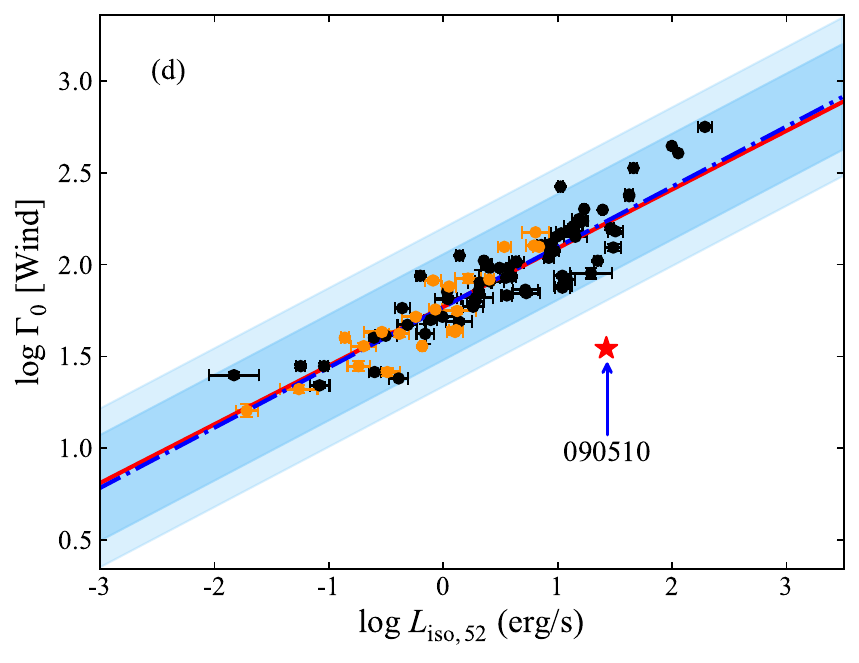}
\includegraphics[angle=0,scale=0.50]{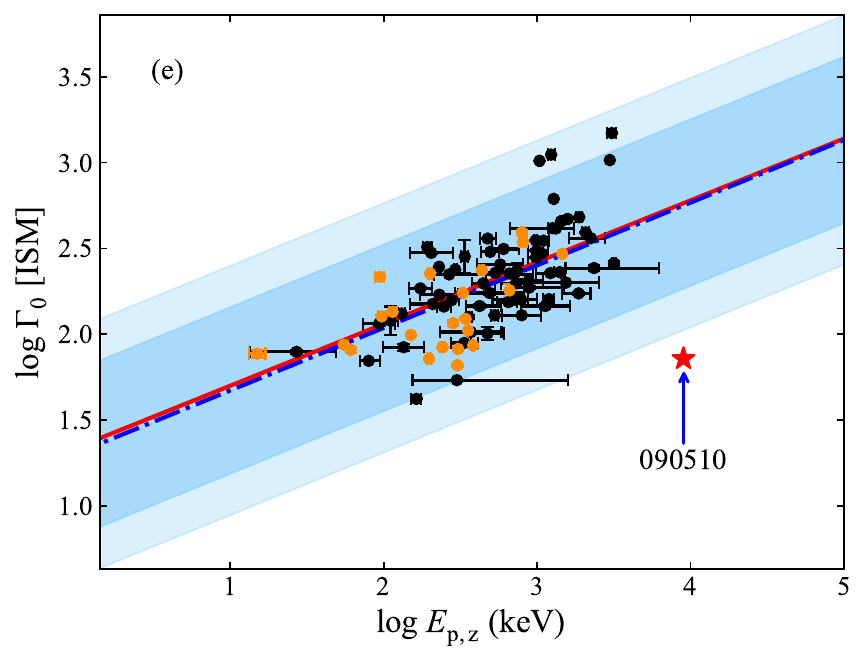}
\includegraphics[angle=0,scale=0.50]{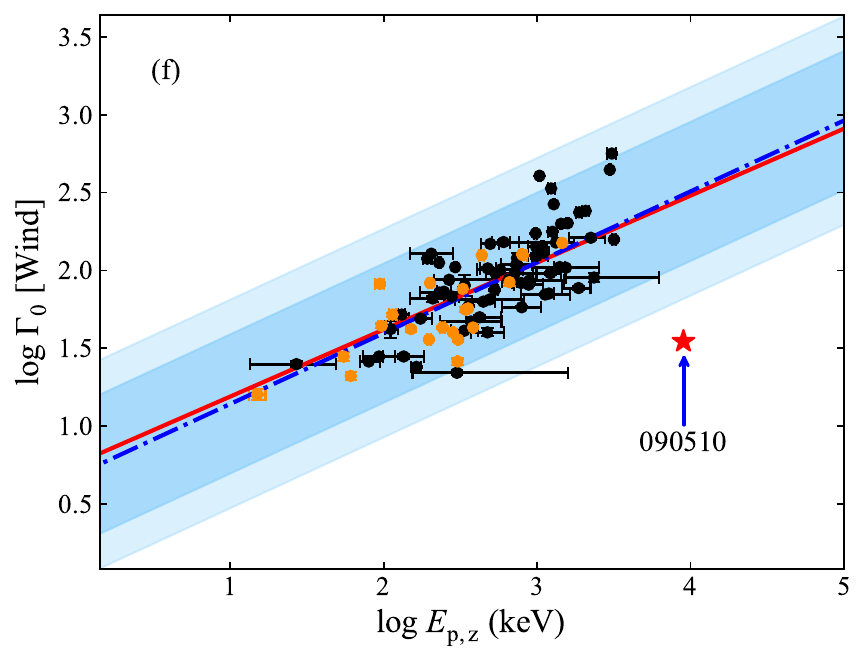}
\caption{Pair relations between $\Gamma_{0}$ and $E_{\rm iso}$ (or $L_{\rm iso}$, $E_{\rm p,z}$) in two different circumburst environments correspond to panels (a)-(f). The orange data points correspond to GRBs whose spectra can only be fitted using the PL model, with their $E_{\rm p}$ values obtained using the method of \citet{2025ApJ...Sun} and \citet{2007ApJ...655L..25Z}. The red solid lines and the blue dotted lines represent the best-fit curves for the 67 GRBs (black data points) with well-measured spectra and the 89 GRBs (all samples), respectively. The blue shaded regions indicate the 2$\sigma$ and 3$\sigma$ prediction level. The red star represents the short GRB 090510.
\label{fig:figure2}}
\end{figure*}

The $\Gamma_{0}$$-$$E_{\rm iso}$ ($L_{\rm iso}$) and $\Gamma_{0}$$-$$E_{\rm p,z}$ relations have been studied by many authors. We re-examine these relations by this large GRBs sample. Figure \ref{fig:figure2} displays the two-parameter correlations between $\Gamma_0$ with $E_{\rm iso}$, $L_{\rm iso}$, and $E_{\rm p,z}$ for 67 GRBs (black data points) with well-measured spectra.
We find that $\Gamma_{0}$ demonstrates a significant dependence on all three quantities. 
We employ the Markov Chain Monte Carlo (MCMC) method \citep{2013PASP..125..306F}\footnote{\url{https://pypi.python.org/pypi/emcee}} to quantify these relations, and the regression results are summarized in Table \ref{tab:tab3}.
The best-fit relations for the $\Gamma_0$–$E_{\rm iso}$ correlation (see Figures \ref{fig:figure2}(a) and (b)) are $\Gamma_0$ (ISM) $\propto$ $E^{0.29 \pm 0.03}_{\rm iso}$, with a Pearson correlation coefficient of $r = 0.74$, a chance probability of $p < 10^{-4}$, and an intrinsic scatter of $\sigma_{\rm int} = 0.21$; and $\Gamma_0$ (Wind) $\propto$ $E^{0.36 \pm 0.02}_{\rm iso}$, with $r$ = 0.90, $p < 10^{-4}$, and $\sigma_{\rm int}$ = 0.14.
The $\Gamma_0$–$L_{\rm iso}$ correlation (see Figures \ref{fig:figure2}(c) and (d)) are $\Gamma_0$ (ISM) $\propto$ $L^{0.28 \pm 0.03}_{\rm iso}$, with $r$ = 0.78, $p < 10^{-4}$, and $\sigma_{\rm int}$ = 0.19; and $\Gamma_0$ (Wind) $\propto$ $L^{0.32 \pm 0.02}_{\rm iso}$, with $r$ = 0.86, $p < 10^{-4}$, and $\sigma_{\rm int}$ = 0.16.
The $\Gamma_{0}$$-$$E_{\rm p,z}$ correlation (see Figures \ref{fig:figure2}(e) and (f)) are $\Gamma_0$ (ISM) $\propto$ $E^{0.36 \pm 0.07}_{\rm p,z}$, with $r$ = 0.53, $p < 10^{-4}$, and $\sigma_{\rm int}$ = 0.26; and $\Gamma_0$ (Wind) $\propto$ $E^{0.43 \pm 0.07}_{\rm p,z}$, with $r$ = 0.63, $p < 10^{-4}$, and $\sigma_{\rm int}$ = 0.24.
As the sample size increases, we find that the correlations between $\Gamma_0$ and $E_{\rm iso}$ (or $L_{\rm iso}$, $E_{\rm p,z}$) remain statistically significant, with tighter relations observed in the wind medium scenario.
This indicates that the initial Lorentz factor of GRBs indeed depends on the radiation energy, luminosity, and peak energy, confirming the findings of previous studies \citep{2010ApJ...725.2209L,2012ApJ...751...49L,2012MNRAS.420..483G}.

\begin{figure*}
\centering
\includegraphics[angle=0,scale=0.50]{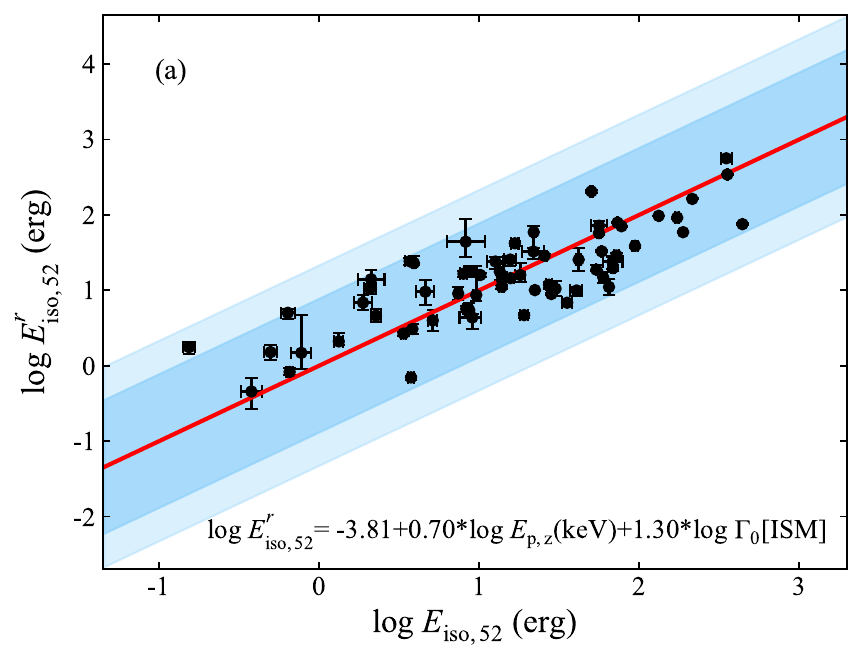}
\includegraphics[angle=0,scale=0.50]{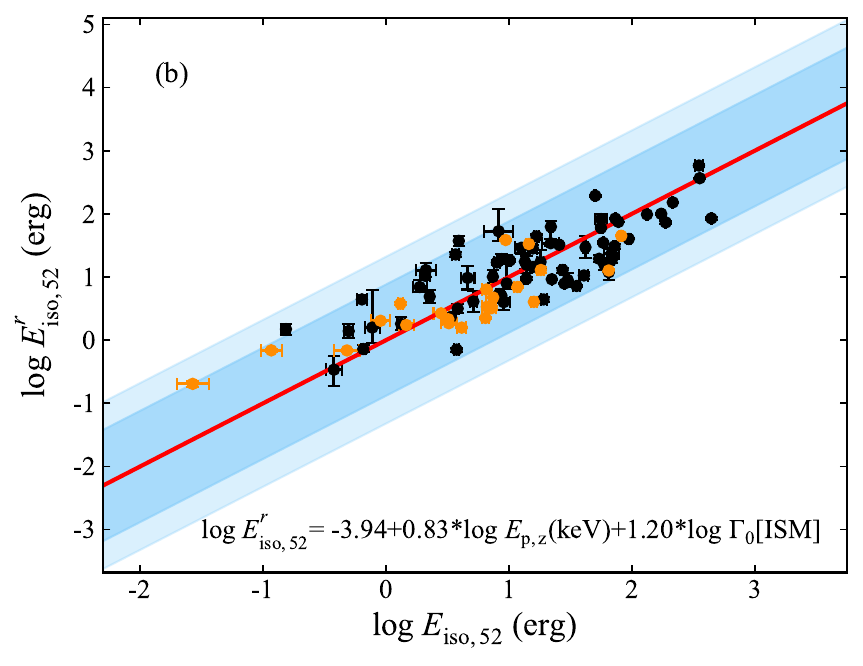}
\includegraphics[angle=0,scale=0.50]{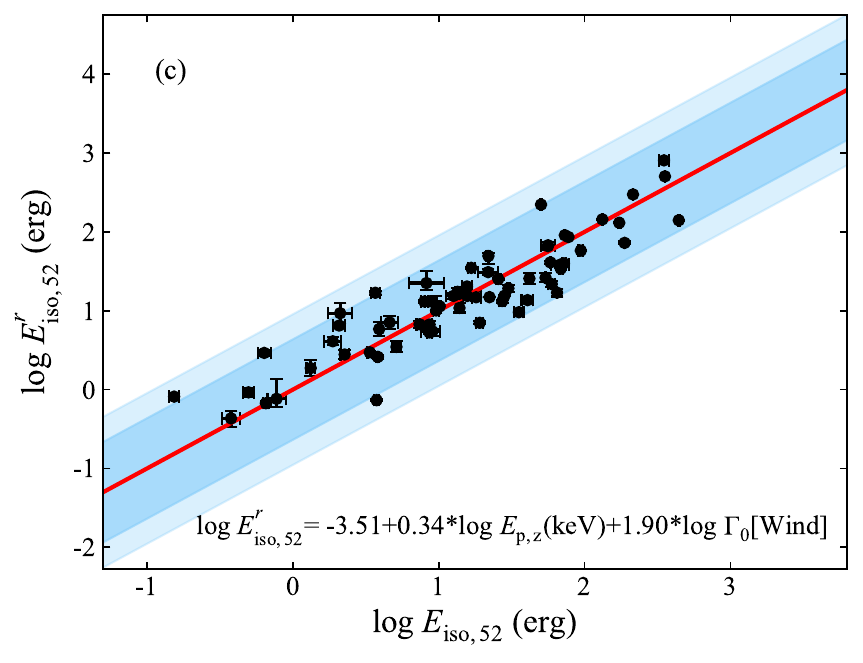}
\includegraphics[angle=0,scale=0.50]{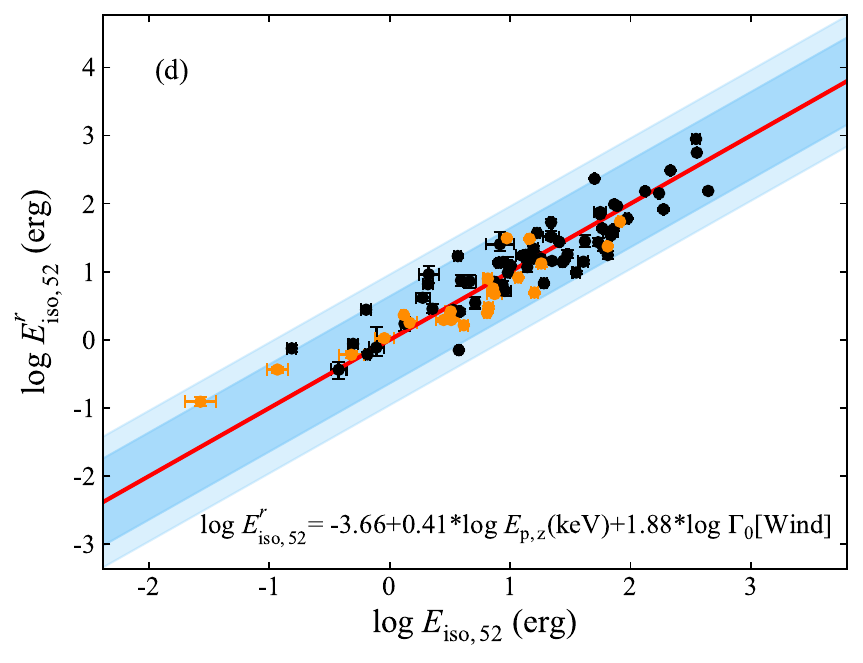}
\includegraphics[angle=0,scale=0.50]{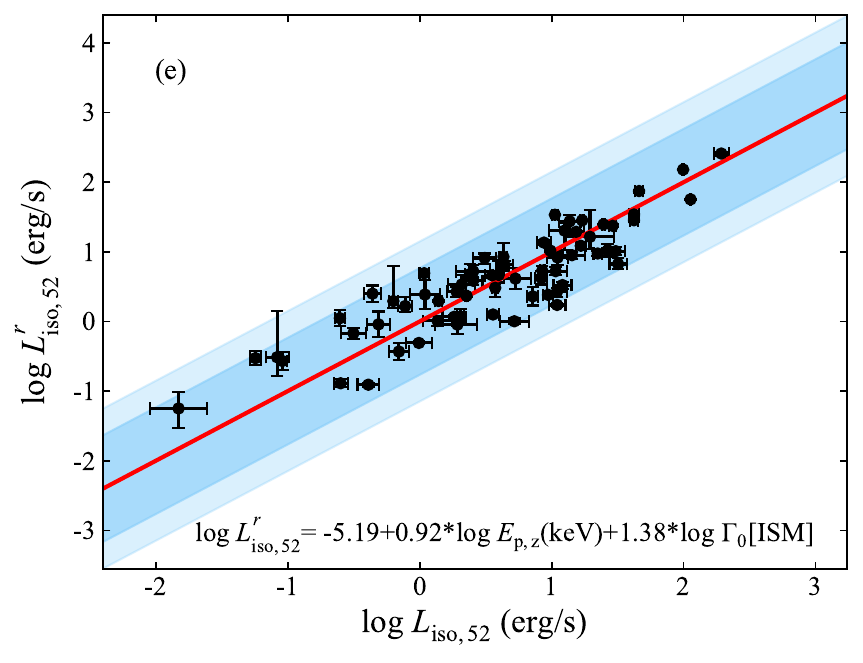}
\includegraphics[angle=0,scale=0.50]{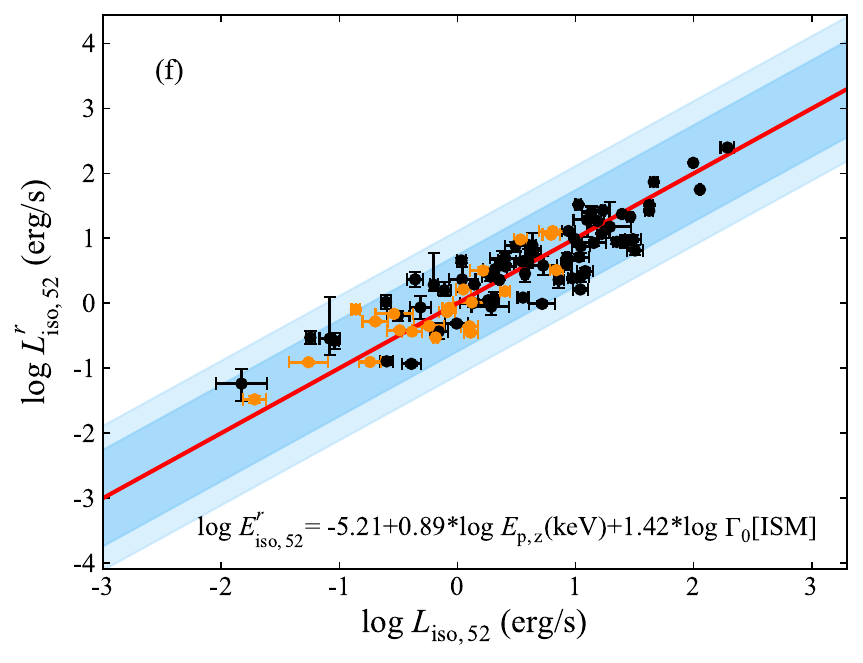}
\includegraphics[angle=0,scale=0.50]{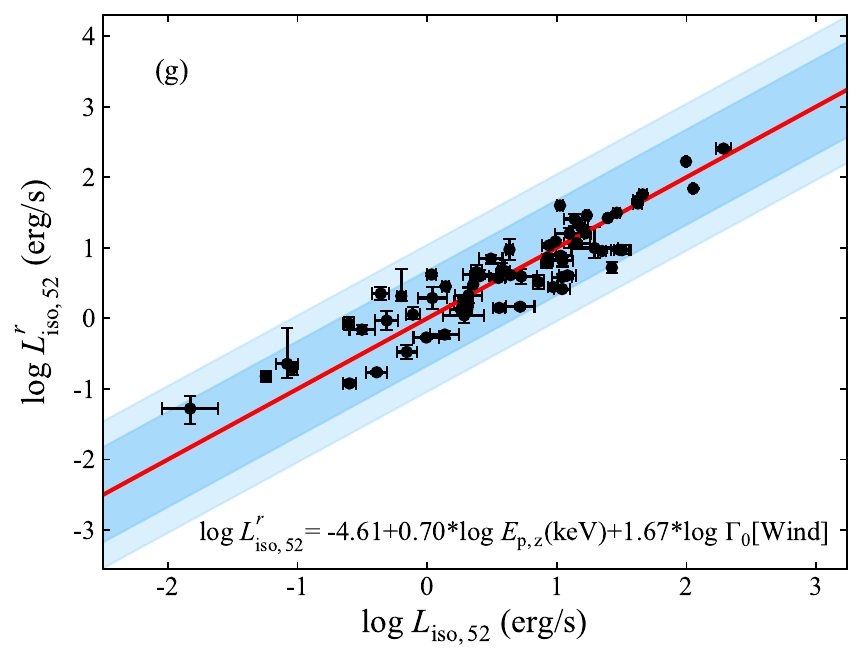}
\includegraphics[angle=0,scale=0.50]{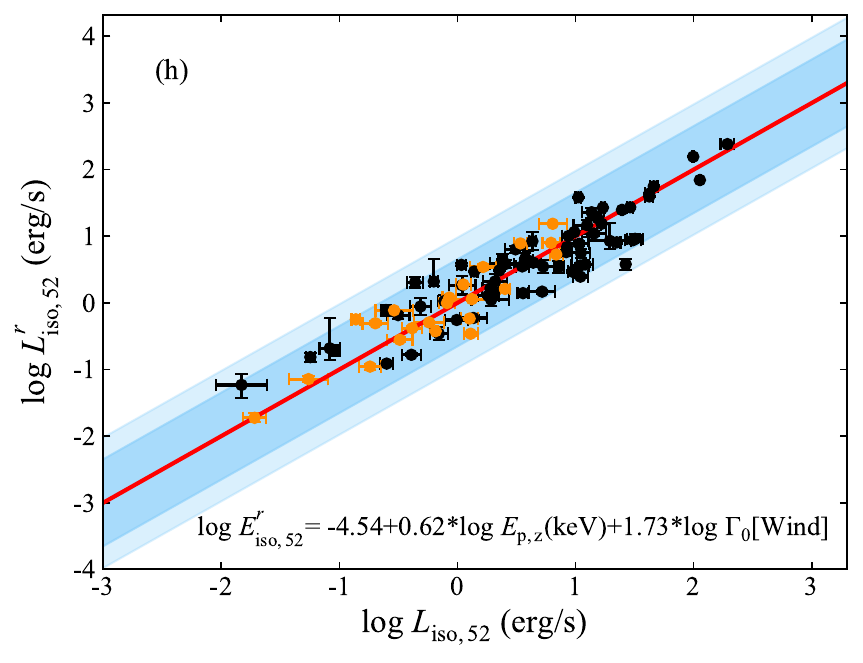}
\caption{The $E^{r}_{\rm iso}$$-$$E_{\rm iso}$ and $L^{r}_{\rm iso}$$-$$L_{\rm iso}$ relations presented for the 67 GRBs with well-measured spectra (left panel a,c,e,g) and all 89 GRBs (right panel b,d,f,h) in two different circumburst environments. The other symbols are the same as in Figure \ref{fig:figure2}.
\label{fig:figure3}}
\end{figure*}

Additionally, we find that the SGRB 090510 significantly deviates from the 3$\sigma$ confidence regions of the $\Gamma_{0}$–$L_{\rm iso}$ and $\Gamma_{0}$–$E_{\rm p,z}$ relations (as shown in Figure \ref{fig:figure2}(c, d, e, f)).
This deviation suggests that intrinsic differences may exist in the central engines or jet launching mechanisms between SGRBs and LGRBs, and that these correlations could potentially be used to distinguish between different GRBs classes.
Previous studies have also reported that SGRB 090510 deviates from the $\Gamma_{0}$-$E_{\rm iso}$ relation \citep{2010ApJ...725.2209L}, although this finding has been challenged by subsequent work \citep{2012ApJ...751...49L,2013ApJ...774...13L}.

In fact, the initial Lorentz factor of this burst remains a topic of ongoing debate.
For instance, the Fermi team provided a lower limit for $\Gamma_{0}$ based on photon opacity constraints ($\Gamma_{\rm 0,min}$ $\sim$ 1200) \citep{2009Natur.462..331A}. 
\citet{2010A&A...510L...7G} assumed that the GeV emission originated from an external shock, leading to an even higher $\Gamma_{0}$ value ($\Gamma_{0} \sim$ 2000)\footnote{Using the peak time derived from the GeV afterglow light curve, our estimates yield $\Gamma_{0} \sim 2078$ (ISM) and $\Gamma_{0} \sim 332$ (Wind), which are broadly consistent with the results of \citet{2010A&A...510L...7G}}.
Later, \citet{2013ApJ...774...13L} estimated a lower $\Gamma_{0}$ value ($\Gamma_{0}$ $\sim$ 61) based on the bump peak observed in the early optical afterglow light curve.
In our work, we also estimate $\Gamma_{0}$ using the optical bump time $t_{\rm p}$, and the results ($\Gamma_{0} = 72$ for the ISM case and $35$ for the wind case) are consistent with those of \citet{2013ApJ...774...13L}.
Due to there is only one SGRB in our sample, the differences between short and long GRBs in the $\Gamma_{0}$$-$$L_{\rm iso}$ and $\Gamma_{0}$$-$$E_{\rm p,z}$ correlations need further investigate in the future.

It is found that the initial Lorentz factor depends not only on the isotropic energy/luminosity but also on the peak energy. 
\citet{2015ApJ...813..116L}  were the first to extend the $\Gamma_{0}$–$E_{\rm iso}$ ($L_{\rm iso}$) relations into three-parameter correlations involving $E_{\rm iso}$ ($L_{\rm iso}$), $E_{\rm p,z}$, and $\Gamma_{0}$ using a sample of 33 GRBs.
More recently, \citet{2023ApJ...949L...4L} and \citet{2023ApJ...952..127Z} conducted systematic studies on the extremely bright GRB 221009A and on several other bright GRBs with detected TeV/sub-TeV gamma-ray afterglows (GRBs 180720B, 190114C, 190829A, 201015A, and 201216C), respectively.
Both studies found that these bursts conform to the $L_{\rm iso}$$-$$E_{\rm p,z}$$-$$\Gamma_{0}$ relation.
We also analyze the correlations between $E_{\rm iso}$ ($E_{\rm p,z}$, $\Gamma_{0}$) and $L_{\rm iso}$ ($E_{\rm p,z}$, $\Gamma_{0}$) for 67 GRBs with well-measured spectra under two different circumburst environment.

The $E_{\rm iso}$$-$$E_{\rm p,z}$$-$$\Gamma_{0}$ correlation is shown in Figures \ref{fig:figure3}(a) and (c), with detailed results provided in Table \ref{tab:tab3}.
The best-fit results are as follows:
\begin{equation}
\begin{aligned} 
E_{\rm iso,52} & = 10^{-3.81 \pm 0.46}  \, (\text{E}_{\rm p,z}/\text{keV})^{0.70 \pm 0.15} \\ & \times \, \Gamma^{1.30 \pm 0.22}_{0}\,(\text{ISM})\, ,  
\label{eq:equation6}
\end{aligned}
\end{equation}
with a Pearson correlation coefficient of $r$ = 0.81, a chance probability of $p < 10^{-4}$, and a dispersion of $\sigma_{\rm int}$ = 0.46; and
\begin{equation}
\begin{aligned} 
E_{\rm iso,52} & = 10^{-3.51 \pm 0.29}  \, (\text{E}_{\rm p,z}/\text{keV})^{0.34 \pm 0.12} \\ & \times \, \Gamma^{1.90 \pm 0.17}_{0}\,(\text{Wind})\, ,  
\label{eq:equation7}
\end{aligned}
\end{equation}
with $r$ = 0.91, $p < 10^{-4}$, and $\sigma_{\rm int}$ = 0.33.
We present $L_{\rm iso}$ as a function of $L_{\rm iso}$ ($E_{\rm p,z}$, $\Gamma_{0}$) in Figures \ref{fig:figure3}(e) and (g), with the regression analysis results also listed in Table \ref{tab:tab3}. 
The best fits are as follows:
\begin{equation}
\begin{aligned} 
L_{\rm iso,52} & = 10^{-5.19 \pm 0.40}  \, (\text{E}_{\rm p,z}/\text{keV})^{0.92 \pm 0.13} \\ 
& \times \, \Gamma^{1.38 \pm 0.19}_{0}\,(\text{ISM})\, ,  
\label{eq:equation8}
\end{aligned}
\end{equation}
with $r$ = 0.89, $p < 10^{-4}$, and $\sigma_{\rm int}$ = 0.39; and
\begin{equation}
\begin{aligned} 
L_{\rm iso,52} & = 10^{-4.61 \pm 0.30}  \, (\text{E}_{\rm p,z}/\text{keV})^{0.70 \pm 0.12} \\ & \times \, \Gamma^{1.67 \pm 0.18}_{0}\,(\text{Wind})\, ,  
\label{eq:equation9}
\end{aligned}
\end{equation}
with $r$ = 0.91, $p < 10^{-4}$, and $\sigma_{\rm int}$ = 0.35.
We find that the $E_{\rm iso}$$-$$E_{\rm p,z}$$-$$\Gamma_0$ and $L_{\rm iso}$$-$$E_{\rm p,z}$$-$$\Gamma_0$ correlations remain significant, although which exhibit relatively large dispersion.

\begin{figure*}
\centering
\includegraphics[angle=0,scale=0.50]{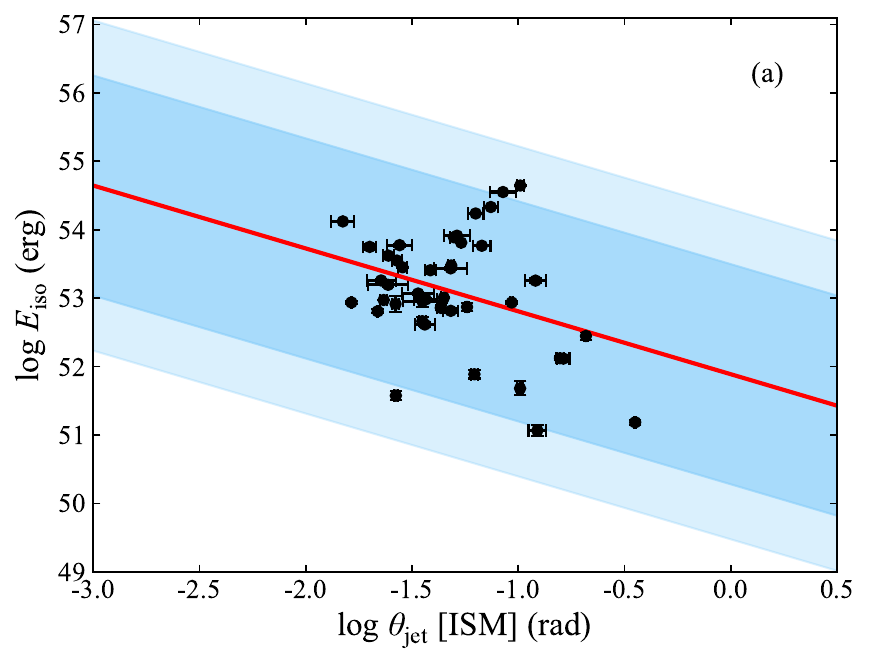}
\includegraphics[angle=0,scale=0.50]{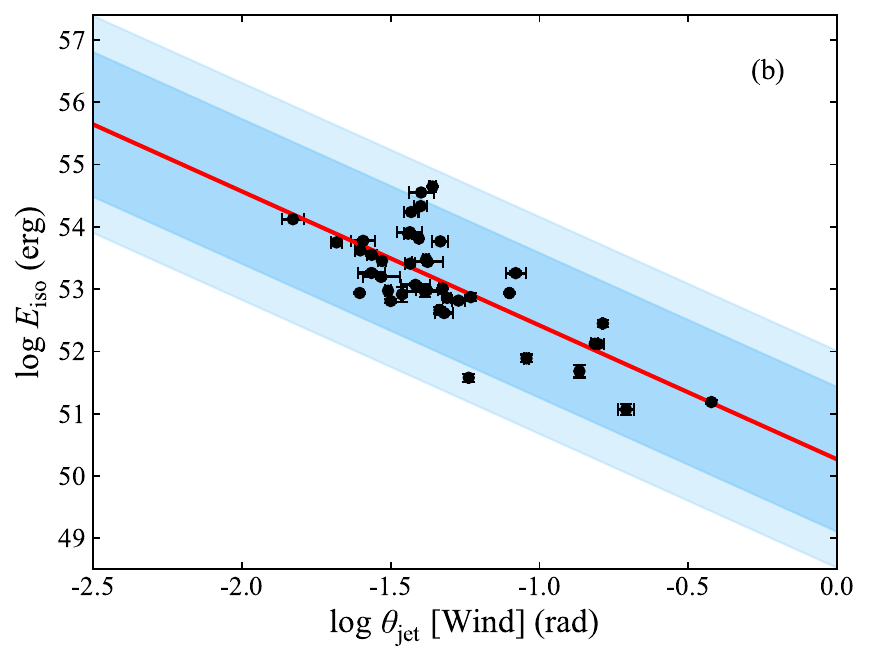}
\includegraphics[angle=0,scale=0.50]{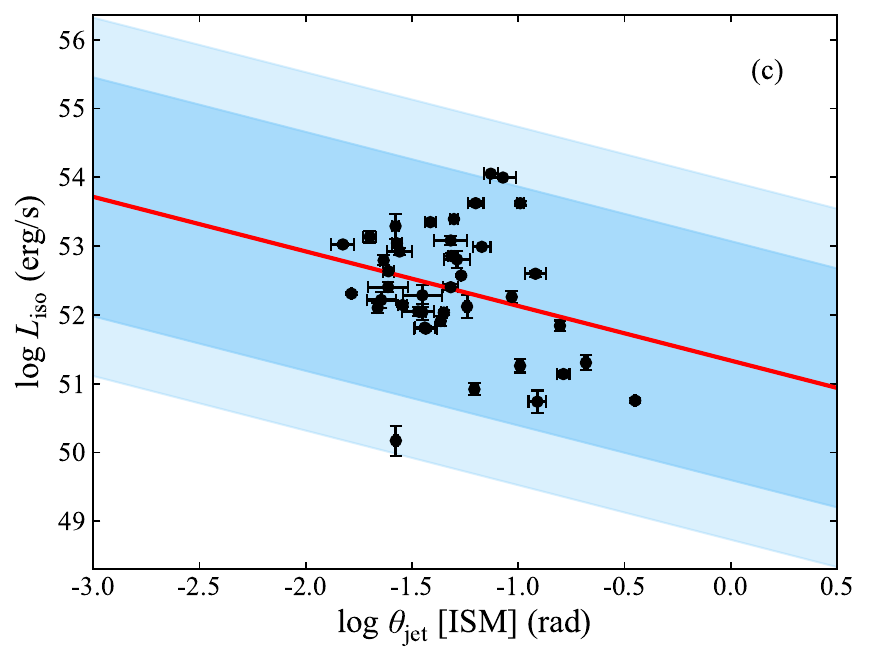}
\includegraphics[angle=0,scale=0.50]{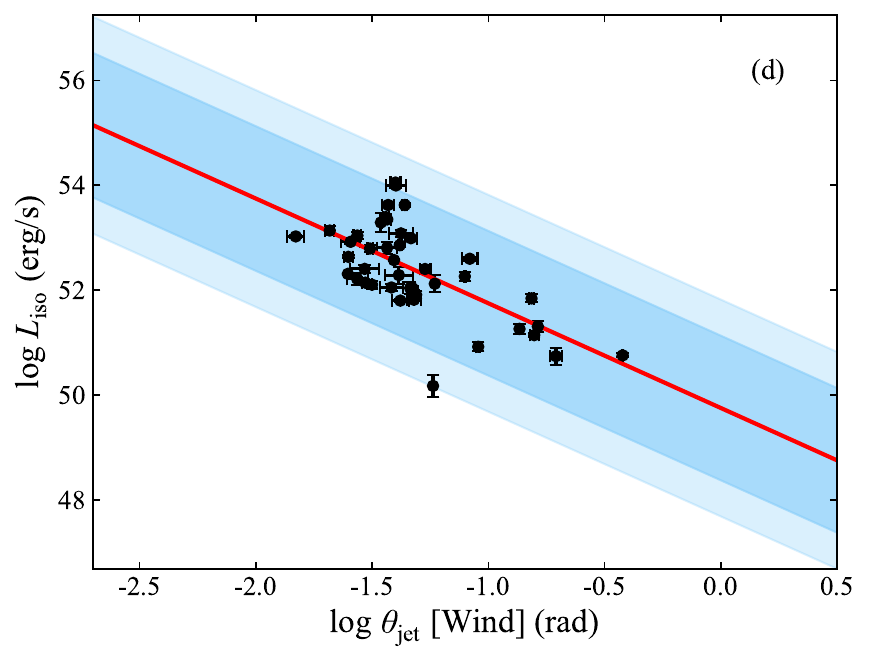}
\includegraphics[angle=0,scale=0.50]{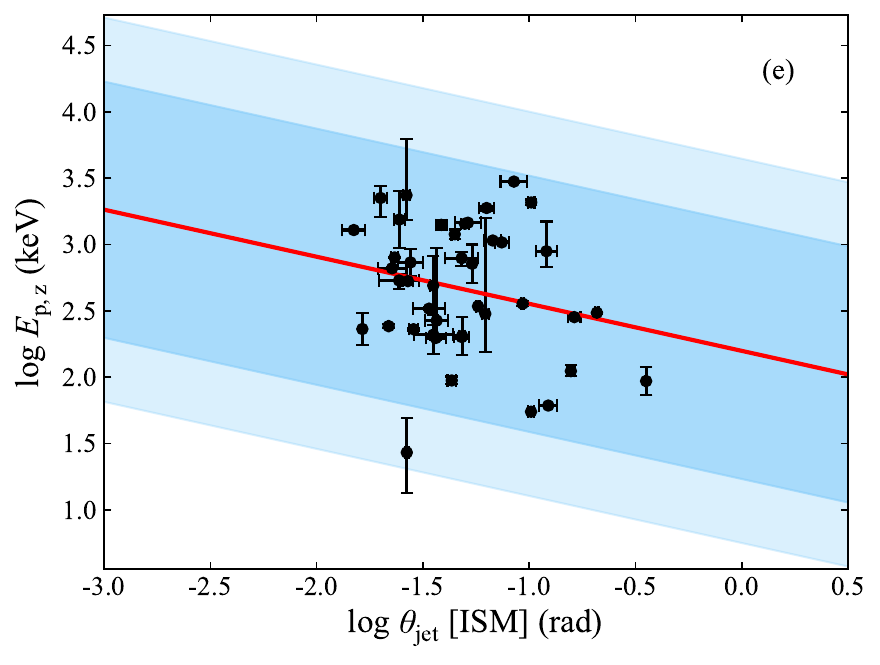}
\includegraphics[angle=0,scale=0.50]{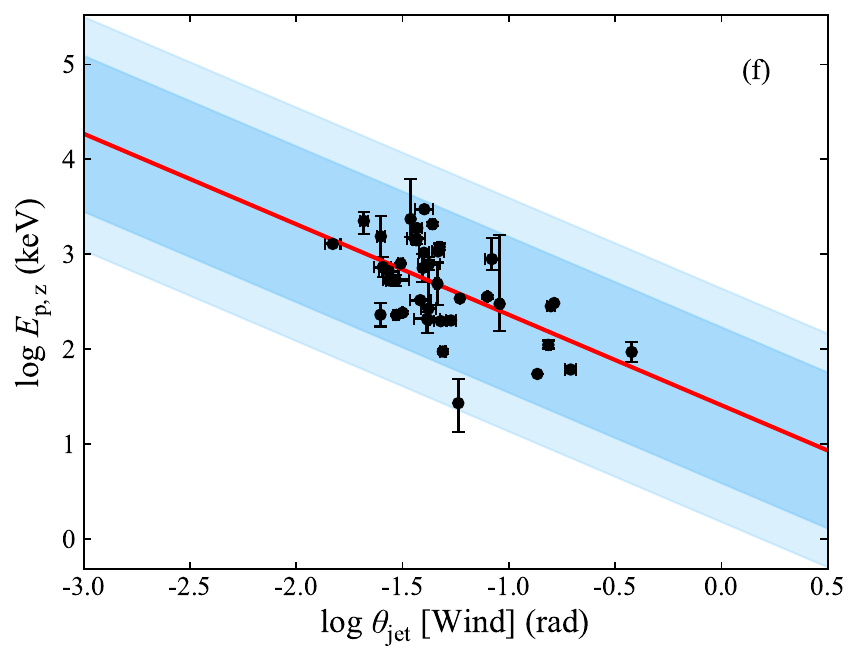}
\caption{Pair relations between $\theta_{\rm jet}$ and $E_{\rm iso}$ (or $L_{\rm iso}$, $E_{\rm p,z}$) in two different circumburst environments correspond to panels (a)-(f). The other symbols are the same as in Figure \ref{fig:figure2}.
\label{fig:figure4}}
\end{figure*}

Moreover, we also explore another form of the three quantities, focusing on $E_{\rm p,z}$ as a function of $L_{\rm iso}$ and $\Gamma_{0}$. The best fits are $E_{\rm p,z}$ $\propto$ $L^{0.50 \pm 0.07}_{\rm iso,52}$$\Gamma^{-0.25 \pm 0.19}_{0}$ (ISM), and $E_{\rm p,z}$ $\propto$ $L^{0.49 \pm 0.09}_{\rm iso,52}$$\Gamma^{-0.21 \pm 0.23}_{0}$ (Wind).
\citet{2002ApJ...581.1236Z} provided a comprehensive overview of various physical models of GRBs and presented the corresponding theoretical predictions for $E_{\rm p,z}$.
For the photosphere model (when the photosphere radius $R_{\rm ph}$ exceeds the coasting radius $R_{\rm c}$), one expects $E_{\rm p,z}$ $\propto$ $L^{-5/12}_{w}$$\Gamma^{8/3}_{0}$ $\propto$ $L^{-5/4}$$\Gamma^{6}_{0}$, where $L_{w}$ is the wind luminosity, $L$ is the photosphere luminosity. Our results clearly do not support the photosphere model. For the synchrotron radiation model in the internal emission region, when the outflow is non-magnetized (i.e., the internal shock model), one expects $E_{\rm p,z}$ $\propto$ $L^{1/2}$$R^{-1}$ $\propto$ $L^{1/2}$ $\Gamma^{-2}_{0}$, where $R$ is the internal shock radius.
The results indicate that the power law index of $L$ in the model presented in this paper agrees with predictions.
However, the index of $\Gamma_{0}$ shows a significant discrepancy, implying that other factors might influence the dependence of $\Gamma_{0}$.

In addition, we perform the same analysis for all 89 GRBs.
The two-parameter and three-parameter correlations are shown in Figures \ref{fig:figure2} and \ref{fig:figure3}(b, d, f, h), respectively, where the best-fit results are summarized in Table \ref{tab:tab3}. 
We find that the power-law indices and Pearson correlation coefficients of the two-parameter and three-parameter relations remain highly consistent with the results from the 67 GRBs with well-measured spectra.
This further confirms that the $E_{\rm iso}$ ($L_{\rm iso}$)$-$$\Gamma_{0}$, $E_{\rm p,z}$$-$$\Gamma_{0}$, and $E_{\rm iso}$ ($L_{\rm iso}$)$-$$E_{\rm p,z}$$-$$\Gamma_{0}$ relations reflect intrinsic physics of GRBs.

\subsection{Correlations of the Jet Half-Opening Angle with the Prompt Emission Parameters and Initial Lorentz Factor} \label{ssec:theta_corre}

\begin{figure*}
\centering
\includegraphics[angle=0,scale=0.80]{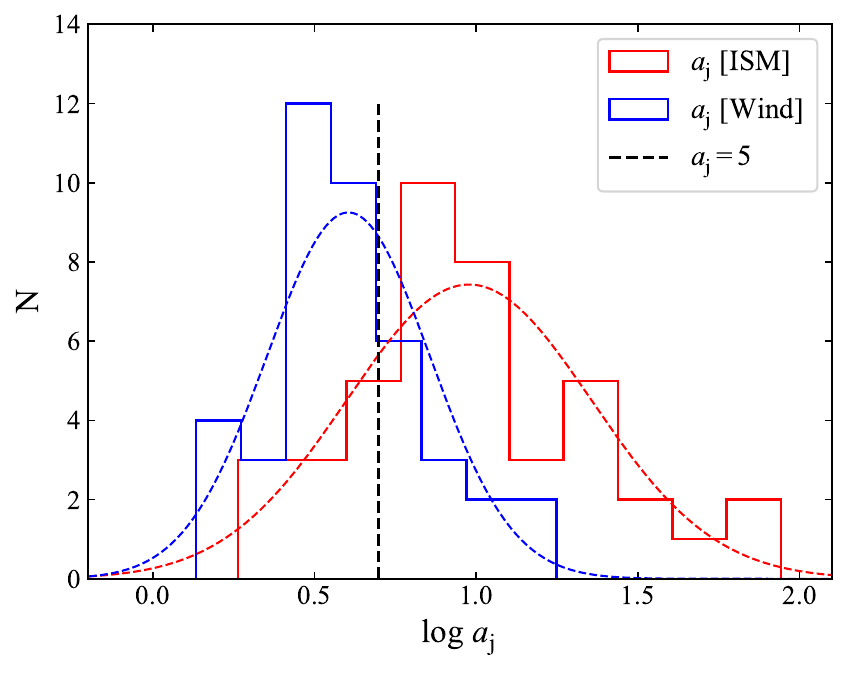}
\caption{Distribution of $a_{\rm j}$. The red and blue lines represent the values of $a_{\rm j}$ for the ISM and wind cases, respectively. The dotted lines are the best-fitting curves. The black vertical dashed line represents the $a_{\rm j}$ value obtained from the empirical relation \citep{2012cosp...39.1354N}.
\label{fig:figure5}}
\end{figure*}

We analyze the two-parameter correlations of $\theta_{\rm jet}$ with $E_{\rm iso}$, $L_{\rm iso}$, and $E_{\rm p,z}$ as shown in Figure \ref{fig:figure4}, and the regression results are provided in Table \ref{tab:tab4}. We find that there is an anti-correlation between $E_{\rm iso}$ and $\theta_{\rm jet}$ (see Figures \ref{fig:figure4}(a) and (b)). 
The best-fits yield $E_{\rm iso}$ $\propto$ $\theta^{-0.92 \pm 0.42}_{\rm jet}$ (ISM), $r$ = -0.34, $p$ = 0.03, and $\sigma_{\rm int}$ = 0.83; and $E_{\rm iso}$ $\propto$ $\theta^{-2.15 \pm 0.33}_{\rm jet}$ (Wind), $r$ = -0.73, $p$ $< 10^{-4}$, and $\sigma_{\rm int}$ = 0.60. 
This results are consistent with the finding of \citet{2020ApJ...900..112Z}, i.e., $E_{\rm iso}$ $\propto$ $\theta^{-0.85 \pm 0.03}_{\rm jet}$ (ISM) and $E_{\rm iso}$ $\propto$ $\theta^{-1.82 \pm 0.03}_{\rm jet}$ (Wind).

From Figure \ref{fig:figure4}, we also obtain weak anti-correlations between $\theta_{\rm jet}$ and both $L_{\rm iso}$ and $E_{\rm p,z}$, although the data exhibit significant dispersion.  
The best-fit results for the $L_{\rm iso}$$-$$\theta_{\rm jet}$ relation (see Figures \ref{fig:figure4}(c) and (d)) are $L_{\rm iso} \propto \theta^{-0.79 \pm 0.45}_{\rm jet}$ (ISM), $r$ = -0.28, $p$ = 0.08, and $\sigma_{\rm int}$ = 0.90; and $L_{\rm iso}$ $\propto$ $\theta^{-2.00 \pm 0.39}_{\rm jet}$ (Wind), $r$ = -0.65, $p$ $< 10^{-4}$, and $\sigma_{\rm int}$ = 0.71.
For the $E_{\rm p,z}$$-$$\theta_{\rm jet}$ relation (see Figures \ref{fig:figure4}(e) and (f)) are $E_{\rm p,z}$ $\propto$ $\theta^{-0.35 \pm 0.25}_{\rm jet}$ (ISM), $r$ = -0.23, $p$ = 0.15, and $\sigma_{\rm int}$ = 0.50; and $E_{\rm p,z}$ $\propto$ $\theta^{-0.95 \pm 0.23}_{\rm jet}$ (Wind), $r$ = -0.56, $p$ = 0.0001, and $\sigma_{\rm int}$ = 0.42.
These results imply that highly energetic and bright GRBs tend to have narrower jet opening angles, signifying that their jets are more collimated.

\citet{2012cosp...39.1354N} proposed an empirical relation to estimate $\theta_{\rm jet}$ based on $\Gamma_0$, i.e., $\theta_{\rm jet} \approx 5.0/ \Gamma_0$.
We tend to test this empirical relation and reformulate it as $a_{\rm j}$ $\approx$ $\Gamma_0$ $\cdot$ $\theta_{\rm jet}$. The distribution of $a_{\rm j}$ is analyzed and shown in Figure \ref{fig:figure5}. We find that the distribution of $a_{\rm j}$ has a large range and is basically log-normal, and the median values of $a_{\rm j} = 9.51^{{}+22.73}_{{}-3.97}$ (ISM) and $a_{\rm j} = 4.03^{{}+7.20}_{{}-2.25}$ (Wind). These results indicate the estimate $\theta_{\rm jet}$ method only based on $\Gamma_0$ proposed by \citet{2012cosp...39.1354N} is not reliable.

\begin{figure*}
\centering
\includegraphics[angle=0,scale=0.50]{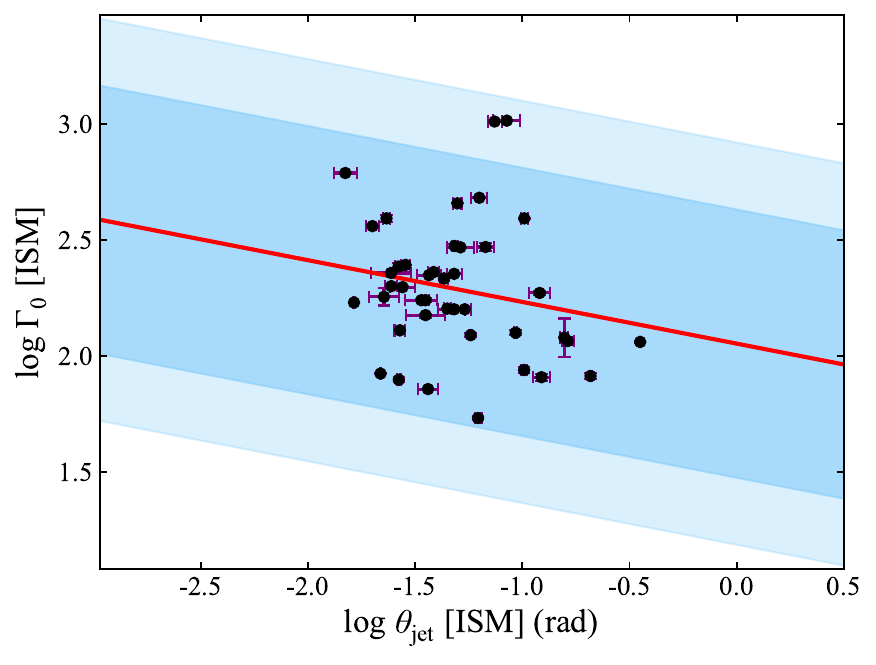}
\includegraphics[angle=0,scale=0.50]{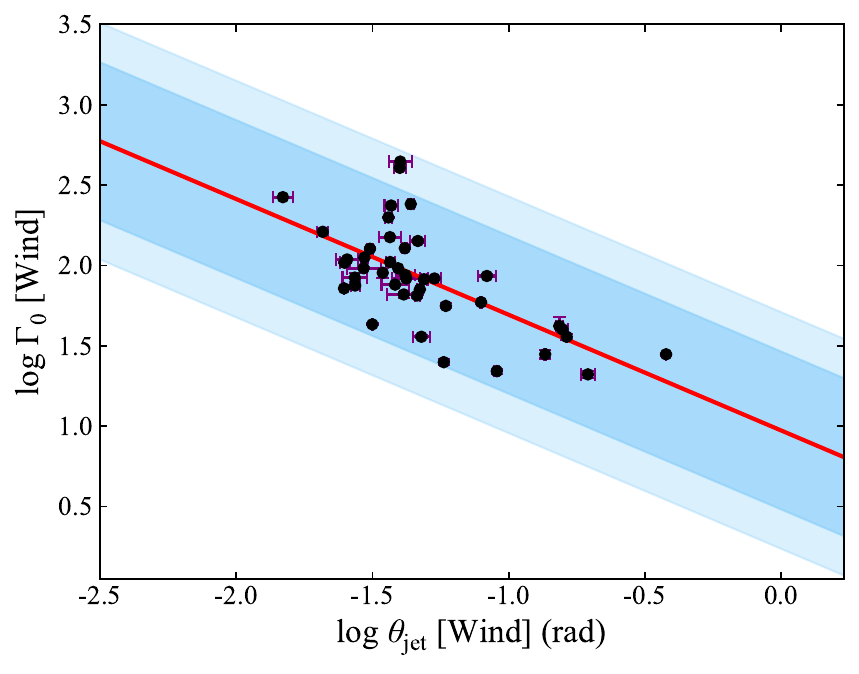}
\caption{Two-parameter relations between $\Gamma_{0}$ and $\theta_{\rm jet}$ in two different circumburst environments. The other symbols are the same as in Figure \ref{fig:figure2}.
\label{fig:figure6}}
\end{figure*}

Based on the $\Gamma_{0}$$-$$E_{\rm iso}$ and $E_{\rm iso}$$-$$\theta_{\rm jet}$ correlations, we speculate that the collimation of GRBs jets may influence the velocity of their relativistic outflows. 
We analyze the relation between $\Gamma_{0}$ and $\theta_{\rm jet}$ and find that $\Gamma_{0}$ indeed depends on $\theta_{\rm jet}$ as shown in Figure \ref{fig:figure6}.
The best-fitting results can be expressed as:
\begin{equation}  
\Gamma_{0}\,(\text{ISM}) = 10^{2.05 \pm 0.20} \theta^{-0.18 \pm 0.15}_{\rm jet}\,(\text{ISM}),    \label{eq:equation10}
\end{equation}
with $r$ = -0.19, $p$ = 0.22, and $\sigma_{\rm int}$ = 0.30; and 
\begin{equation}  
\Gamma_{0}\,(\text{Wind}) = 10^{0.97 \pm 0.19} \theta^{-0.72 \pm 0.14}_{\rm jet}\,(\text{Wind}),    \label{eq:equation11}
\end{equation}
with $r$ = -0.65, $p < 10^{-4}$, and $\sigma_{\rm int}$ = 0.25.
These relations indicate that GRBs with more collimated jets tend to have higher initial Lorentz factors, implying faster outflow velocities.
\citet{2013MNRAS.428.1410G} extensively discussed these two key parameters using numerical simulation methods.
They found that larger values of $\Gamma_{0}$ (i.e. the ‘faster’ bursts) correspond to smaller values of $\theta_{\rm jet}$ (i.e. the ‘narrower’). The numerical simulation result is consistent with our finding.

\begin{figure*}
\centering
\includegraphics[angle=0,scale=0.50]{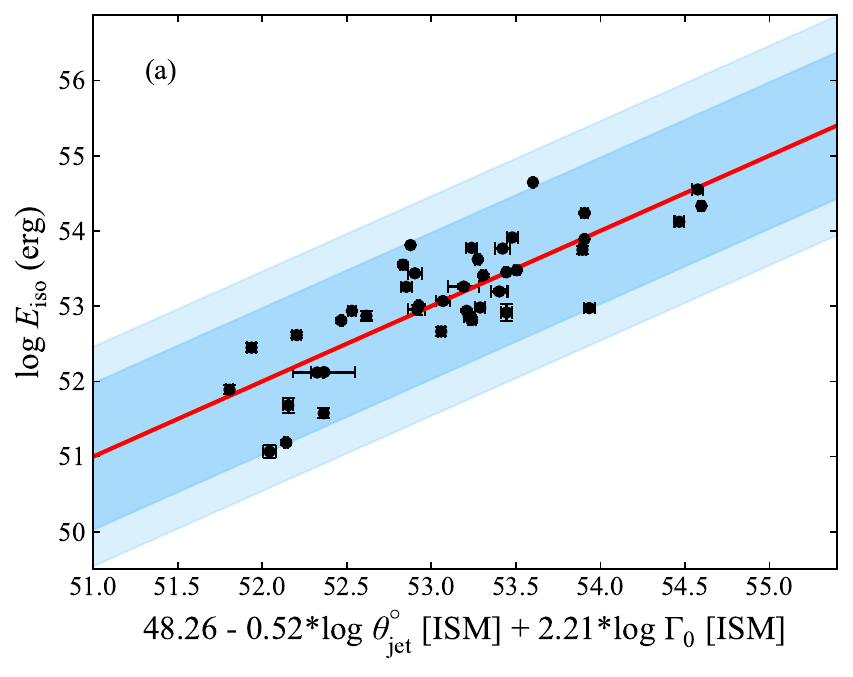}
\includegraphics[angle=0,scale=0.50]{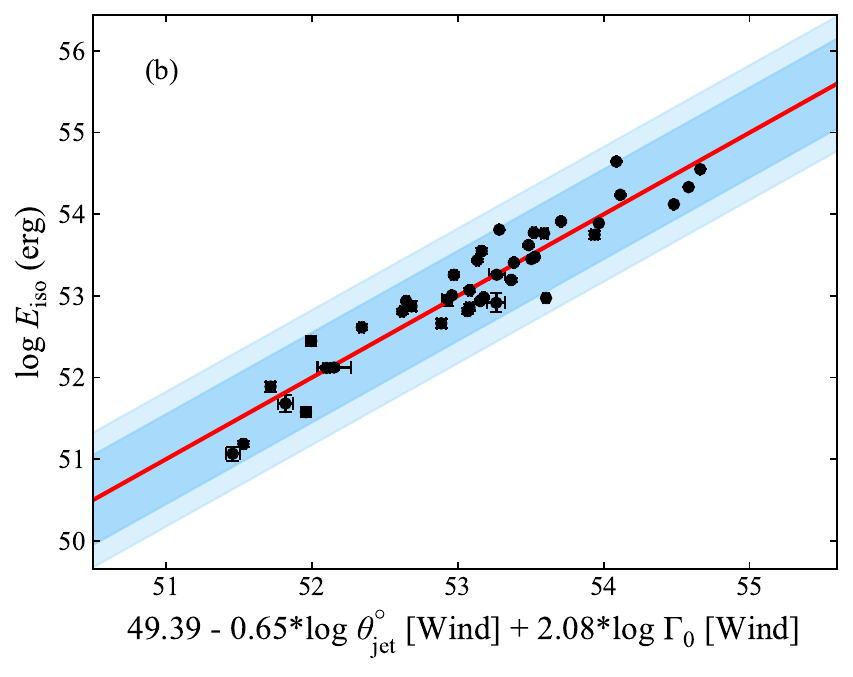}
\includegraphics[angle=0,scale=0.50]{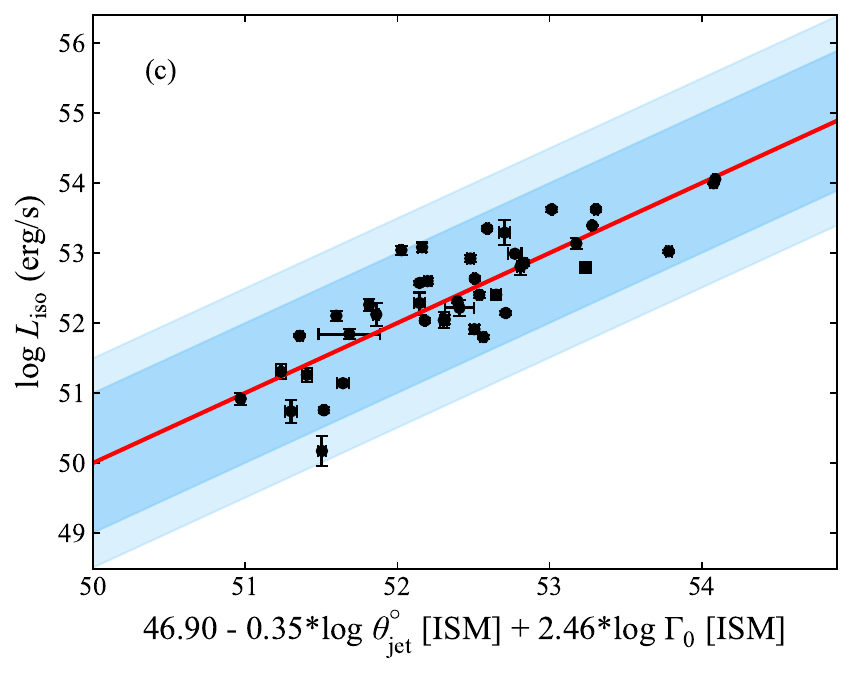}
\includegraphics[angle=0,scale=0.50]{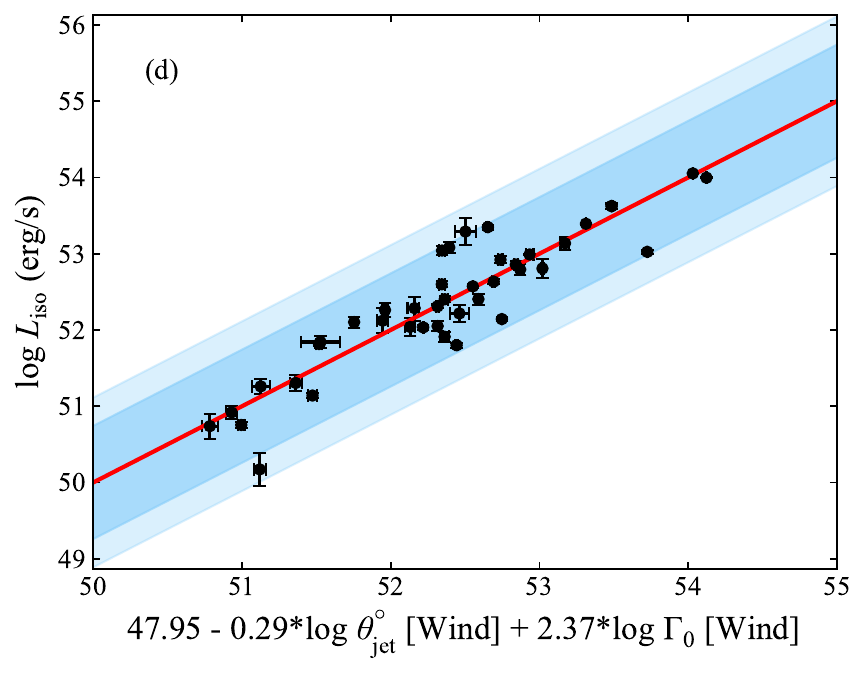}
\includegraphics[angle=0,scale=0.50]{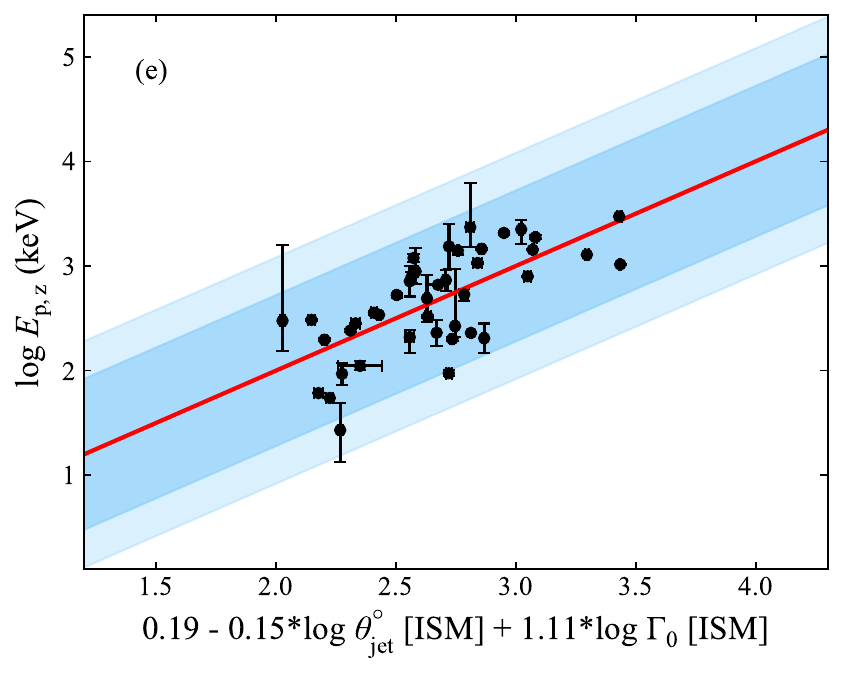}
\includegraphics[angle=0,scale=0.50]{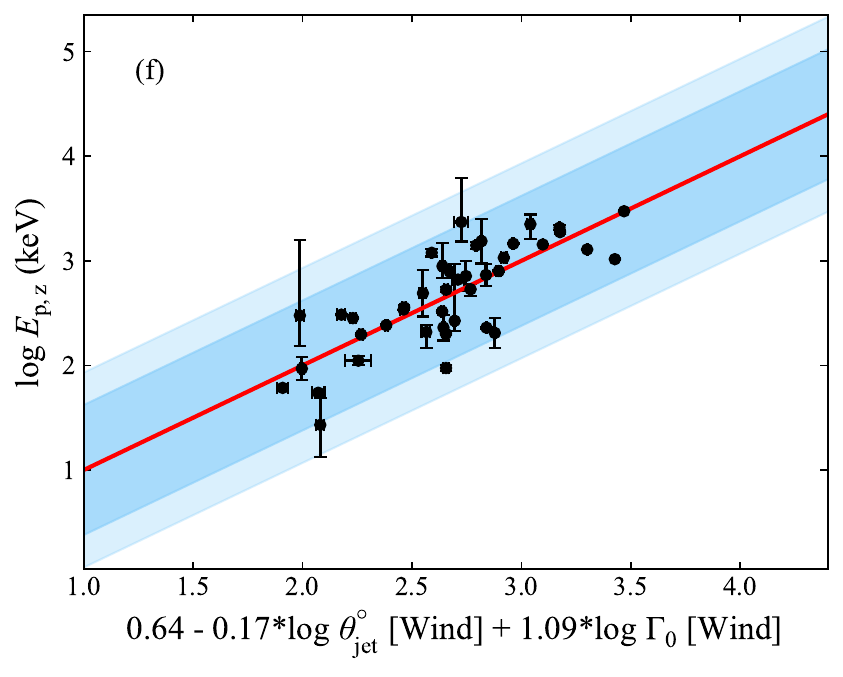}
\caption{Three new relations among $E_{\rm iso}$ (or $L_{\rm iso}$, $E_{\rm p,z}$), $\theta_{\rm jet}$, and $\Gamma_0$ are presented across two different circumburst environments in panels (a)-(f). The other symbols are the same as in Figure \ref{fig:figure2}.
\label{fig:figure7}}
\end{figure*}

In addition, we further explore other potential three-parameter correlations among $E_{\rm iso}$ (or $L_{\rm iso}$, $E_{\rm p,z}$), $\Gamma_0$, and $\theta_{\rm jet}$.
Figure \ref{fig:figure7} shows the three-parameter correlations derived from the variable sets \{$E_{\rm iso}$, $\theta_{\rm jet}$, $\Gamma_0$\}, \{$L_{\rm iso}$, $\theta_{\rm jet}$, $\Gamma_0$\} and \{$E_{\rm p,z}$, $\theta_{\rm jet}$, $\Gamma_0$\}. The best-fit results are summarized in Table \ref{tab:tab4}.
We find three remarkably tight three-parameter correlations: the $E_{\rm iso}$$-$$\theta_{\rm jet}$$-$$\Gamma_0$, $L_{\rm iso}$$-$$\theta_{\rm jet}$$-$$\Gamma_0$ and $E_{\rm p,z}$$-$$\theta_{\rm jet}$$-$$\Gamma_0$ relations.

The best-fitting results for the $E_{\rm iso}$$-$$\theta_{\rm jet}$$-$$\Gamma_0$ correlation (see Figures \ref{fig:figure7}(a) and (b)) are given by:
\begin{equation}
\begin{aligned} 
E_{\rm iso} & = 10^{48.26 \pm 0.68}  \theta^{-0.52 \pm 0.26}_{\rm jet}\,(\text{ISM}) \\
& \times \Gamma^{2.21 \pm 0.28}_{0}\,(\text{ISM}) ,   
\label{eq:equation12}
\end{aligned}
\end{equation}
with $r$ = 0.82, $p < 10^{-4}$,  and $\sigma_{\rm int}$ = 0.51; and
\begin{equation}
\begin{aligned} 
E_{\rm iso} & = 10^{49.39 \pm 0.43} \theta^{-0.65 \pm 0.21}_{\rm jet}\,(\text{Wind}) \\
& \times \Gamma^{2.08 \pm 0.19}_{0}\,(\text{Wind}) ,   
\label{eq:equation13}
\end{aligned}
\end{equation}
with $r$ = 0.95, $p < 10^{-4}$,  and $\sigma_{\rm int}$ = 0.29.
The $L_{\rm iso}$$-$$\theta_{\rm jet}$$-$$\Gamma_0$ relation (see Figures \ref{fig:figure7}(c) and (d)) are expressed as:
\begin{equation}
\begin{aligned} 
L_{\rm iso} & = 10^{46.90 \pm 0.70} \theta^{-0.35 \pm 0.27}_{\rm jet}\,(\text{ISM}) \\
& \times \Gamma^{2.46 \pm 0.29}_{0}\,(\text{ISM}) ,  
\label{eq:equation14}
\end{aligned}
\end{equation}
with $r$ = 0.83, $p < 10^{-4}$,  and $\sigma_{\rm int}$ = 0.52; and
\begin{equation}
\begin{aligned} 
L_{\rm iso} & = 10^{47.95 \pm 0.57} \theta^{-0.29 \pm 0.28}_{\rm jet}\,(\text{ISM}) \\
& \times \Gamma^{2.37 \pm 0.25}_{0}\,(\text{ISM}) ,  
\label{eq:equation15}
\end{aligned}
\end{equation}
with $r$ = 0.91, $p < 10^{-4}$,  and $\sigma_{\rm int}$ = 0.39.
Similarly, the $E_{\rm p,z}$$-$$\theta_{\rm jet}$$-$$\Gamma_0$ correlation (see Figures \ref{fig:figure7}(e) and (f)) are:
\begin{equation}
\begin{aligned} 
E_{\rm p,z} & = 10^{0.19 \pm 0.51} \theta^{-0.15 \pm 0.19}_{\rm jet}\,(\text{ISM}) \\
& \times \Gamma^{1.11 \pm 0.21}_{0}\,(\text{ISM}) ,   
\label{eq:equation16}
\end{aligned}
\end{equation}
with $r$ = 0.69, $p < 10^{-4}$,  and $\sigma_{\rm int}$ = 0.38; and
\begin{equation}
\begin{aligned} 
E_{\rm p,z} & = 10^{0.64 \pm 0.48} \theta^{-0.17 \pm 0.23}_{\rm jet}\,(\text{Wind}) \\
& \times \Gamma^{1.09 \pm 0.21}_{0}\,(\text{Wind}) ,   
\label{eq:equation17}
\end{aligned}
\end{equation}
with $r$ = 0.78, $p < 10^{-4}$,  and $\sigma_{\rm int}$ = 0.32.

\subsection{Correlations between the Initial Lorentz Factor with beaming-corrected Prompt Emission Parameters} \label{ssec:E_gamma_corre}

\begin{figure*}
\centering
\includegraphics[angle=0,scale=0.50]{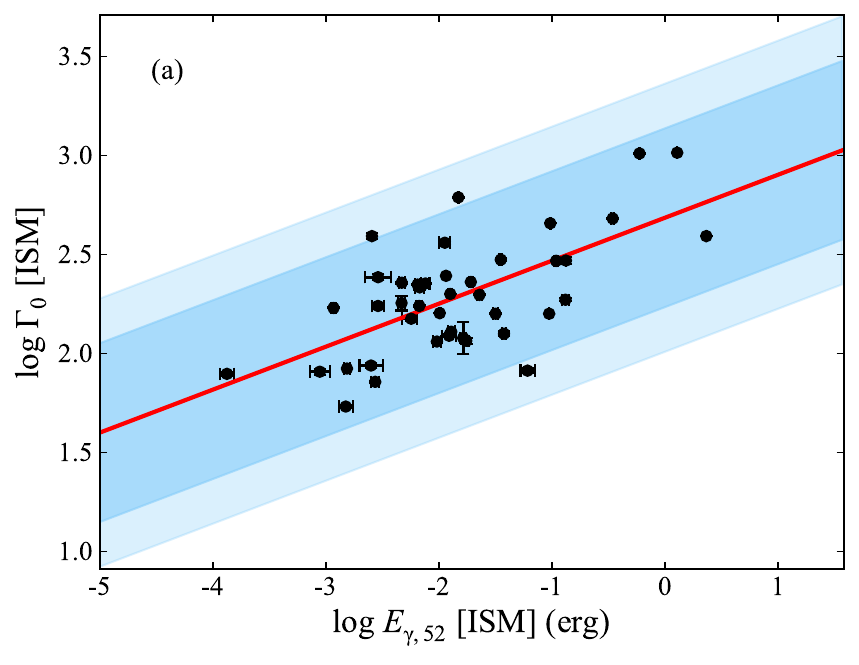}
\includegraphics[angle=0,scale=0.50]{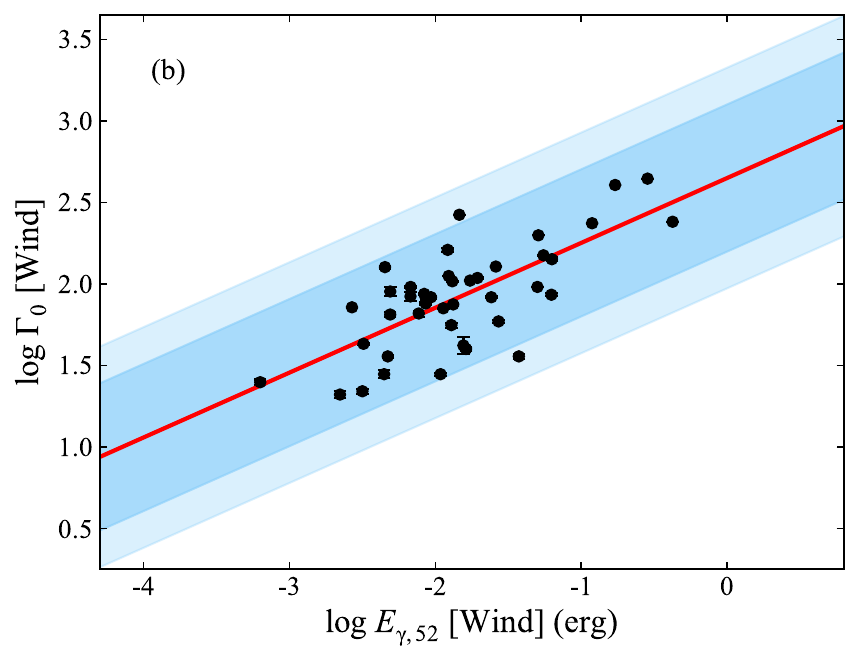}
\includegraphics[angle=0,scale=0.50]{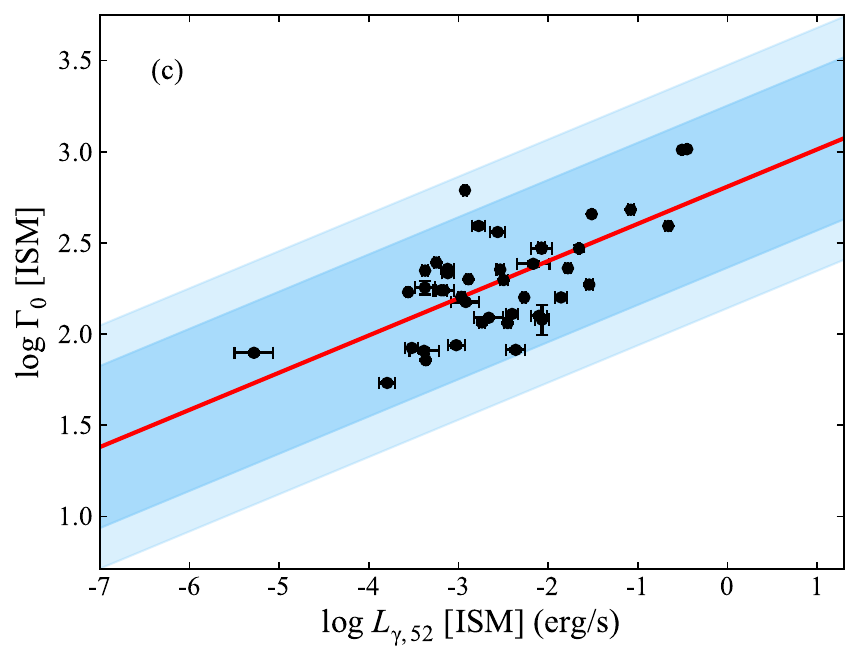}
\includegraphics[angle=0,scale=0.50]{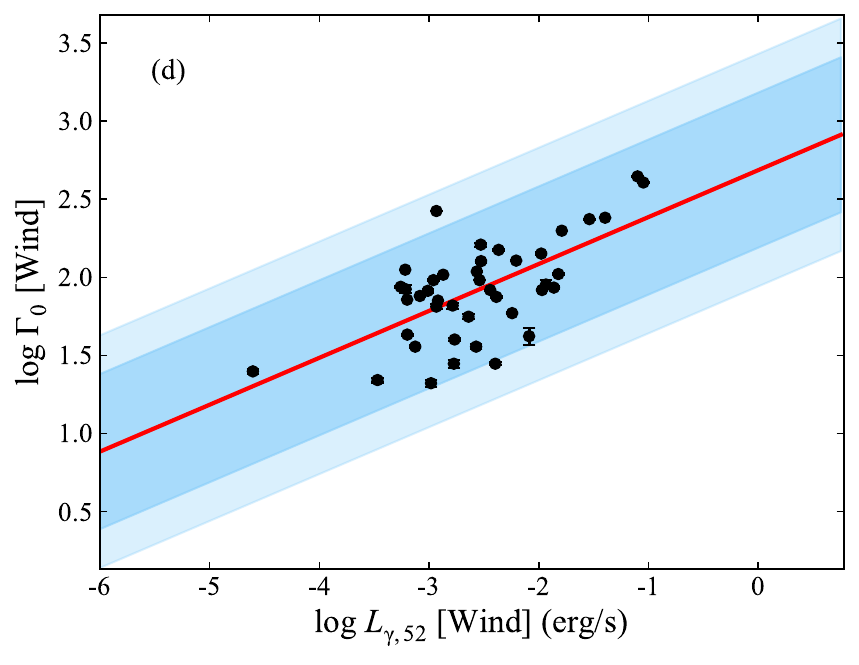}
\caption{The beaming-corrected two-parameter relations between $\Gamma_{\rm 0}$ and $E_{\rm \gamma}$ (or $L_{\rm \gamma}$) in two different circumburst environments correspond to panels (a)-(d). The other symbols are the same as in Figure \ref{fig:figure2}.
\label{fig:figure8}}
\end{figure*}

Using the values of the opening angel of GRBs, we can further analyze the relations between the beaming-corrected energy, luminosity, and other physical parameters of GRBs. The jet-corrected isotropic energy and peak luminosity are defined as follows:
\begin{equation}    
E_{\gamma} = E_{\text{iso}} \cdot f_{\text{b}},  \label{eq:equation18}  
\end{equation}  
\begin{equation}    
L_{\rm \gamma} = L_{\text{iso}} \cdot f_{\text{b}}, \label{eq:equation19}  
\end{equation}
where $f_{\rm b}$ is the beaming correction factor, i.e.,
\begin{equation}  
f_{\rm b} = (1 - \cos\theta_{\rm jet}) \, .
\label{eq:equation20}
\end{equation}
The beaming-corrected energy $E_{\rm \gamma,52}$ and luminosity $L_{\gamma,52}$ values are also listed in Table \ref{tab:tab2}.

We re-examine the correlations between $\Gamma_{0}$ and $E_{\gamma}$ /$L_{\gamma}$ (see Figure \ref{fig:figure8}).
The results of the regression analysis are presented in Table \ref{tab:tab5}.
For the $E_{\gamma}$$-$$\Gamma_{0}$ relation (see Figures \ref{fig:figure8}(a) and (b)), the best-fit results are given by: 
\begin{equation}  
\Gamma_{0}\,(\text{ISM}) = 10^{2.69 \pm 0.09} E^{0.22 \pm 0.04}_{\gamma,52} \, (\text{ISM}) \,, \label{eq:equation21}
\end{equation}
with $r$ = 0.64, $p < 10^{-4}$, and $\sigma_{\rm int}$ = 0.23; and 
\begin{equation}  
\Gamma_{0}\,(\text{Wind}) = 10^{2.65 \pm 0.12} E^{0.40 \pm 0.06}_{\gamma,52}\, (\text{Wind}) \,,  \label{eq:equation22}
\end{equation}
withe $r$ = 0.72, $p < 10^{-4}$, and $\sigma_{\rm int}$ = 0.23.
For the $L_{\gamma}$$-$$\Gamma_{0}$ relation (see Figures \ref{fig:figure8}(c) and (d)), the best fits yield:
\begin{equation}  
\Gamma_{0}\,(\text{ISM}) = 10^{2.81 \pm 0.10} L^{0.20 \pm 0.04}_{\gamma,52}\,(\text{ISM}),   \label{eq:equation23}
\end{equation}
with $r$ = 0.66, $p < 10^{-4}$, and $\sigma_{\rm int}$ = 0.23; and 
\begin{equation}  
\Gamma_{0}\,(\text{Wind}) = 10^{2.69 \pm 0.16} L^{0.30 \pm 0.06}_{\gamma,52}\,(\text{Wind}),  \label{eq:equation24}
\end{equation}
with $r$ = 0.64, $p < 10^{-4}$, and $\sigma_{\rm int}$ = 0.26.
From the above results, it is evident that the correlations between the beaming-corrected $E_{\gamma}$ (or $L_{\gamma}$) and $\Gamma_{0}$ remain statistically significant.
However, compared to the pre-correction relations, both relations exhibit greater dispersion. 

Theoretically, in the fireball shock model, the relativistic jet dissipates kinetic energy through internal shocks and produces gamma-ray emission.
For simplicity, we assume a constant radiation efficiency $\eta$ for each burst, leading to $L_{\gamma}$ $=$ $\eta$$\dot{E}_{\nu\bar{\nu}}$, where $\dot{E}_{\nu\bar{\nu}}$ is the neutrino annihilation power.
Based on simulations of a black hole central engine, \citet{2013ApJ...765..125L} derived the relation $\Gamma_{0} \propto \dot{E}^{0.27}_{\nu\bar{\nu}}$; consequently, we obtain $\Gamma_{0} \propto L^{0.27}_{\gamma}$.
It can be seen that the statistical correlations for both the ISM and wind profiles are consistent with the expected results of the black hole model proposed by \citet{2013ApJ...765..125L}.
\citet{2017JHEAp..13....1Y} also analyzed these correlations using a sample of 17 GRBs with observed jet breaks, and found that the strongly magnetized Blandford–Znajek (BZ) mechanism \citep{1977MNRAS.179..433B} of a black hole central engine is more consistent with their observational results.
These findings further support the possibility that black holes serve as candidate central engines of GRBs.

\begin{figure*}
\centering
\includegraphics[angle=0,scale=0.50]{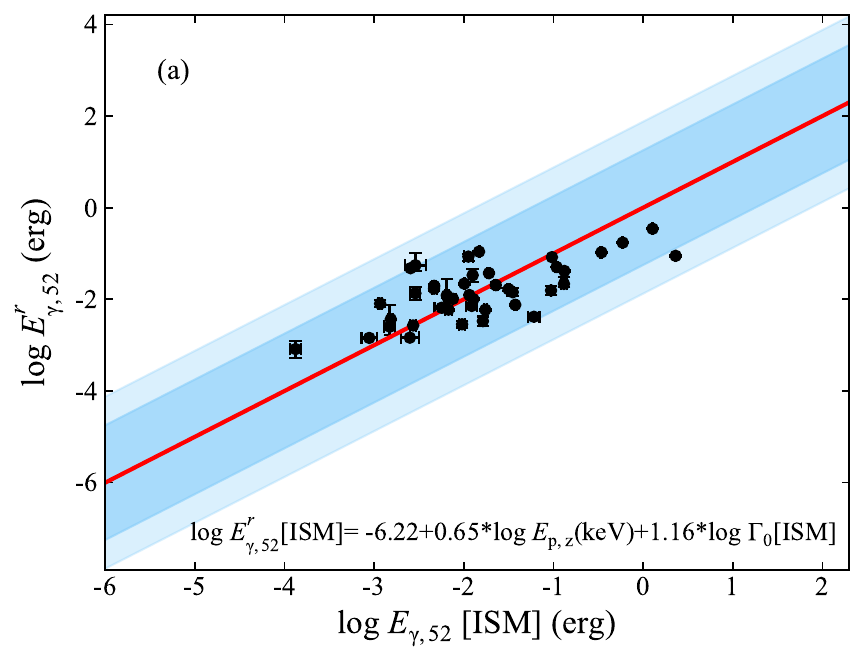}
\includegraphics[angle=0,scale=0.50]{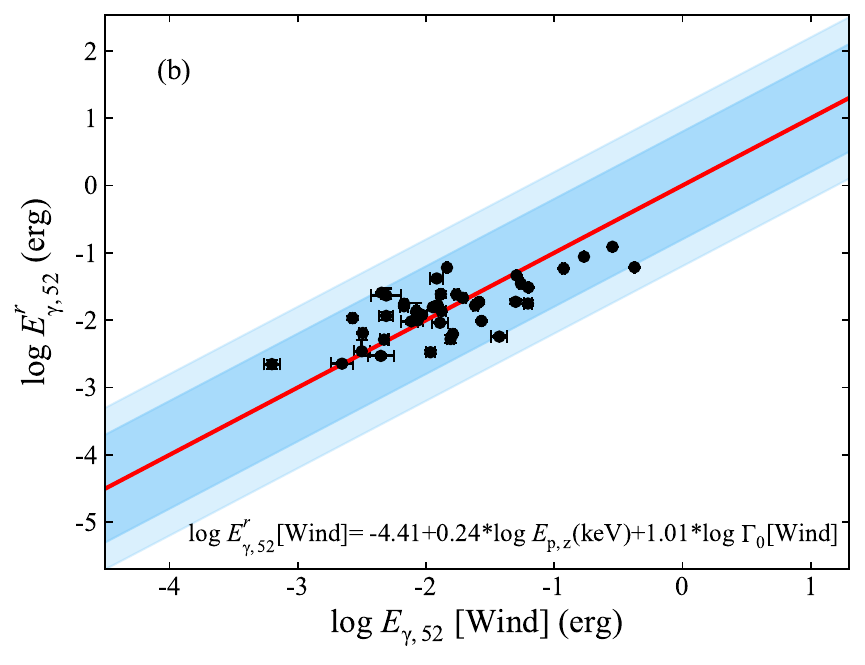}
\includegraphics[angle=0,scale=0.50]{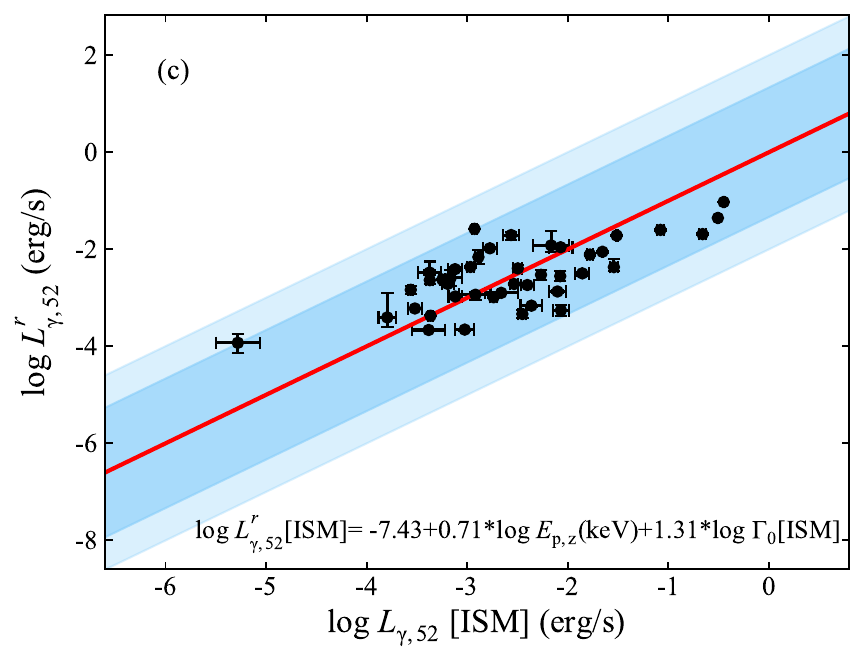}
\includegraphics[angle=0,scale=0.50]{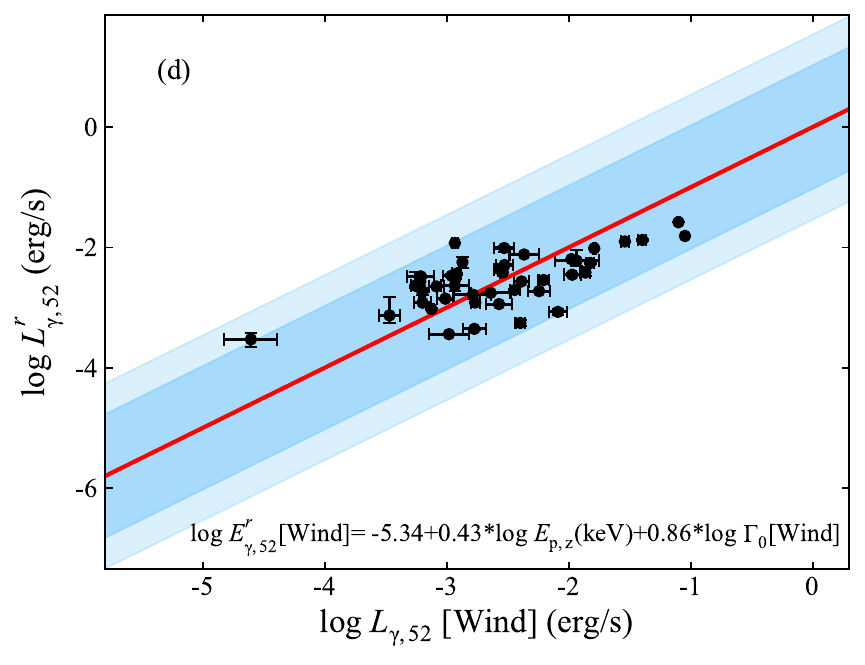}
\caption{The beaming-corrected relations of $E^{r}_{\rm \gamma}$$-$$E_{\rm \gamma}$ and $L^{r}_{\rm \gamma}$$-$$L_{\rm \gamma}$ in two different circumburst environments correspond to panels (a)-(d). The other symbols are the same as in Figure \ref{fig:figure2}.
\label{fig:figure9}}
\end{figure*}

Additionally, we explore the $E_{\gamma}$$-$$E_{\rm p,z}$$-$$\Gamma_{0}$ and $L_{\gamma}$$-$$E_{\rm p,z}$$-$$\Gamma_{0}$ correlations and find that these three-parameter relations are indeed exist in our sample, as shown in Figure \ref{fig:figure9}.
The detailed fitting results are listed in Table \ref{tab:tab5}.
The best-fit fits for the $E_{\gamma}$$-$$E_{\rm p,z}$$-$$\Gamma_{0}$ correlation (see Figures \ref{fig:figure9}(a) and (b)) are given by:
\begin{equation}
\begin{aligned} 
E_{\gamma,52}\,(\text{ISM}) & = 10^{-6.22 \pm 0.82} (\text{E}_{\rm p,z}/\text{keV})^{0.65 \pm 0.29} \\ & \times \Gamma^{1.16 \pm 0.48}_{0}\,(\text{ISM}) \, ,  
\label{eq:equation25}
\end{aligned}
\end{equation}
with $r$ = 0.70, $p < 10^{-4}$, and $\sigma_{\rm int}$ = 0.65; and 
\begin{equation}
\begin{aligned}  
E_{\gamma,52}\,(\text{Wind}) & = 10^{-4.41 \pm 0.41}(\text{E}_{\rm p,z}/\text{keV})^{0.24 \pm 0.21} \\ & \times \Gamma^{1.01 \pm 0.33}_{0}\,(\text{Wind}) \, ,
\label{eq:equation26}
\end{aligned}
\end{equation}
with $r$ = 0.73, $p < 10^{-4}$, and $\sigma_{\rm int}$ = 0.42.
Similarly, the best-fit relations for the $L_{\gamma}$$-$$E_{p,z}$$-$$\Gamma_{0}$ correlation (see Figures \ref{fig:figure8}(c) and (d)) are:
\begin{equation}
\begin{aligned} 
L_{\gamma,52}\,(\text{ISM}) & = 10^{-7.43 \pm 0.87} (\text{E}_{\rm p,z}/\text{keV})^{0.71 \pm 0.31} \\ & \times \Gamma^{1.31 \pm 0.52}_{0}\,(\text{ISM}) \, ,  
\label{eq:equation27}
\end{aligned}
\end{equation}
with $r$ = 0.71, $p < 10^{-4}$, and $\sigma_{\rm int}$ = 0.70; and 
\begin{equation}
\begin{aligned} 
L_{\gamma,52}\,(\text{Wind}) & = 10^{-5.34 \pm 0.52}  (\text{E}_{\rm p,z}/\text{keV})^{0.43 \pm 0.27} \\ & \times \Gamma^{0.86 \pm 0.42}_{0}\,(\text{Wind}) \, ,  
\label{eq:equation28}
\end{aligned}
\end{equation}
with $r$ = 0.67, $p < 10^{-4}$, and $\sigma_{\rm int}$ = 0.54.
We find that the relations all become weaker and the dispersion increases after jet correction.

\section{Conclusion} \label{sec:concl}
We compile the largest sample to data of GRBs with well measured early afterglow onset bump time $t_{\rm p}$, totaling 89 events, among which 42 display jet break features.
Where the $t_{\rm p}$ values of 81 GRBs (80 LGRBs and 1 SGRB) are determined from early optical afterglow light curves. While for the remaining 8 GRBs, which are extracted from GeV light curves.
Based on early bump peak time and jet break time, we estimate the initial Lorentz factors and jet half-opening angles for the sample.
We perform a comprehensive analysis of the correlations between these two key physical parameters and the prompt emission parameters, and discuss the possible underlying physical nature. 
The main results are summarized below:
\begin{enumerate}
    \item For the 89 GRBs sample with onset bumps, the initial Lorentz factor $\Gamma_{0}$ spans a wide range, from several tens to over a thousand. The median values under two different medium cases are $\Gamma_{0} =$ $174^{{}+335}_{{}-90}$ (ISM) and $\Gamma_{0} = $ $80^{{}+160}_{{}-40}$ (Wind). Among the 42 GRBs with jet breaks, the jet half-opening angle $\theta_{\rm jet}$ shows similar distributions in both ISM and wind media, with median values of $\theta_{\rm jet} = $ $2.80^{{}+5.70}_{{}-1.38}$\textdegree\,(ISM) and $\theta_{\rm jet} = $ $2.78^{{}+5.35}_{{}-1.44}$\textdegree\,(Wind).

    \item Using this large sample with measured $\Gamma_{0}$, we confirm the previously reported correlations of $\Gamma_{0}$$-$$E_{\rm iso}$ ($L_{\rm iso}$), $\Gamma_{0}$$-$$E_{\rm p,z}$, and $E_{\rm iso}$ ($L_{\rm iso}$)$-$$E_{\rm p,z}$$-$$\Gamma_{0}$ in both ISM and wind medium cases. We also find that SGRB 090510 significantly deviates from the 3$\sigma$ confidence range of the $\Gamma_{0}$$-$$L_{\rm iso}$ and $\Gamma_{0}$$-$$E_{\rm p,z}$ relations. As this is the only short GRB in our sample, whether its deviation reflects a genuinely different distribution requires further investigation with more sample. Moreover, compared with the predictions of the photospheric model, the $E_{\rm p,z}$$-$$L_{\rm iso}$$-$$\Gamma_{0}$ relation provides stronger support for the synchrotron radiation model.

    \item By analyzing 42 GRBs with jet breaks, we investigate the relations between the jet opening angle and the prompt emission. We further confirm the $\theta_{\rm jet}$$-$$E_{\rm iso}$ correlation and find weak anti-correlations between $\theta_{\rm jet}$ and both $L_{\rm iso}$ and $E_{\rm p,z}$, i.e., $L_{\rm iso} \propto \theta^{-0.79 \pm 0.45}_{\rm jet}$\,(ISM) and $L_{\rm iso} \propto \theta^{-2.00 \pm 0.39}_{\rm jet}$\,(Wind); $E_{\rm p,z} \propto \theta^{-0.35 \pm 0.25}_{\rm jet}$\,(ISM) and $E_{\rm p,z} \propto \theta^{-0.95 \pm 0.23}_{\rm jet}$\,(Wind). Furthermore, we explore the relation between the jet opening angle and the initial Lorentz factor. For the first time, we report a dependence of $\Gamma_{0}$ on $\theta_{\rm jet}$, i.e., $\Gamma_{0} \propto \theta^{-0.18 \pm 0.15}_{\rm jet}$\,(ISM) and $\Gamma_{0} \propto \theta^{-0.72 \pm 0.14}_{\rm jet}$\,(Wind). 
    This results indicate highly energetic and bright GRBs tend to be more collimated and have higher velocities.
    Additionally, we perform a detailed analysis of the multi-parameter correlations among \{$E_{\rm iso}$, $\theta_{\rm jet}$, $\Gamma_0$\}, \{$L_{\rm iso}$, $\theta_{\rm jet}$, $\Gamma_0$\} and \{$E_{\rm p,z}$, $\theta_{\rm jet}$, $\Gamma_0$\}, and, for the fist time, find three new tight three-parameter $E_{\rm iso}$$-$$\theta_{\rm jet}$$-$$\Gamma_{0}$, $L_{\rm iso}$$-$$\theta_{\rm jet}$$-$$\Gamma_{0}$ and $E_{\rm p,z}$$-$$\theta_{\rm jet}$$-$$\Gamma_{0}$ correlations. 
    
    \item Following beaming corrections to both the radiated energy and luminosity, we update the relations between the initial Lorentz factor and the corrected energy, i.e., $\Gamma_{0}$\,(\rm ISM) $\propto$ $E^{0.22\pm0.04}_{\gamma,52}$\,(ISM) and
    $\Gamma_{0}$\,(Wind) $\propto$
    $E^{0.40\pm0.06}_{\gamma,52}$\,(Wind); $\Gamma_{0}$\,(ISM) $\propto$ $L^{0.20\pm0.04}_{\gamma,52}$\,(ISM) and $\Gamma_{0}$\,(Wind) $\propto$ $L^{0.30\pm0.06}_{\gamma,52}$\,(\text{Wind}). We find that the $L_{\gamma}$$-$$\Gamma_{0}$ relations for both the ISM and wind medium cases show excellent agreement with the predictions of a black hole central engine powered by neutrino–antineutrino annihilation. Meanwhile, when taking $E_{\rm p,z}$ into account, the three-parameter $E_{\gamma}$ ($L_{\gamma}$)$-$$E_{\rm p,z}$$-$$\Gamma_{0}$ relations remain valid. However, compared to the pre-correction $E_{\rm iso}$ ($L_{\rm iso}$)$-$$E_{\rm p,z}$$-$$\Gamma_{0}$ relations, these corrected correlations are significantly weakened, and the dispersion increases notably.
\end{enumerate}

\section*{Acknowledgments}
We thank the anonymous referee for insightful comments/suggestions.
We also acknowledge the use of public data from the Swift , Fermi and Konus-Wind catalogs.
We thank Si-Yuan Zhu for the helpful discussions.
This work was supported in part by the National Natural Science Foundation of China (grant No. 12463008), and by the Guangxi Natural Science Foundation (grant No. 2022GXNSFDA035083).
L.L.Z. acknowledges the support by the Foundation of Guilin University of Technology (No. RD2500000384).

\bibliography{GRB_gamma0}

\clearpage

\begin{longrotatetable}
\setlength{\tabcolsep}{10pt}
\begin{deluxetable*}{lcccccccc}
\tablecaption{The Parameters of GRBs with Onset Bump Features
\label{tab:tab1}}
\tablewidth{750pt}
\tabletypesize{\scriptsize}
\tabletypesize{\normalsize}
\tablehead{
\multicolumn{1}{l}{GRBs} & \colhead{$z^{a}$} & \colhead{$E^{b}_{\rm p,z}$} & \colhead{$E_{\rm iso,52}$} & \colhead{$L_{\rm iso,52}$}& \colhead{log $t^{c}_{\rm p}$} & \colhead{$\Gamma_{0}$} & \colhead{$\Gamma_{0}$} & \colhead{Ref.} \\
\colhead{} & \colhead{} & \colhead{(keV)} & \colhead{(erg)}  & \colhead{(erg/s)}& \colhead{(s)} & \colhead{(ISM)} & \colhead{(Wind)} & \colhead{$t_{\rm p}$}   
}

\startdata
990123     & 1.6     & $   1882.40^{+80.60}_{-88.40}$ & $    172.78^{+4.34}_{-4.45}$ & $     42.09^{+3.33}_{-3.33}$ &   1.68 &$  481.00^{+1.51}_{-1.55}$ & $  236.00^{+1.48}_{-1.52}$ & (1) \\
050502A    & 3.793   & $    797.86^{+9.77}_{-9.77}$$^\diamond$ & $      9.40^{+0.00}_{-0.00}$ & $      6.21^{+1.04}_{-1.04}$ &   1.76 &$  392.00^{+12.66}_{-12.66}$ & $  127.00^{+2.73}_{-2.73}$ & (2) \\
050730     & 3.967   & $    661.31^{+10.62}_{-10.62}$$^\diamond$ & $     18.16^{+1.16}_{-1.16}$ & $      1.65^{+0.42}_{-0.42}$ &   2.78 &$  180.00^{+15.65}_{-15.65}$ & $   84.00^{+5.04}_{-5.04}$ & (3) \\
050820A    & 2.612   & $    889.00^{+459.00}_{-239.00}$$^\star$ & $     18.06^{+1.27}_{-1.27}$ & $      3.97^{+0.37}_{-0.37}$ &   2.59 &$  187.00^{+3.41}_{-3.41}$ & $   86.00^{+1.77}_{-1.77}$ & (3) \\
050904     & 6.29    & $   1458.41^{+17.80}_{-17.80}$$^\diamond$ & $     81.87^{+3.05}_{-3.05}$ & $      6.40^{+1.76}_{-1.76}$ &   2.59 &$  294.00^{+1.37}_{-1.37}$ & $  150.00^{+1.40}_{-1.40}$ & (4) \\
050922C    & 2.198   & $   2350.53^{+2292.97}_{-1007.37}$ & $      8.24^{+2.24}_{-2.24}$ & $     19.52^{+8.13}_{-8.13}$ &   2.12 &$  243.00^{+8.25}_{-8.25}$ & $   90.00^{+6.12}_{-6.12}$ & (1) \\
060124     & 2.296   & $    787.74^{+85.70}_{-112.06}$ & $     27.29^{+1.23}_{-1.23}$ & $     12.07^{+1.90}_{-1.90}$ &   2.80 &$  159.00^{+0.90}_{-0.90}$ & $   83.00^{+0.94}_{-0.94}$ & (1) \\
060210     & 3.91    & $    732.00^{+172.00}_{-172.00}$$^\star$ & $     59.52^{+3.18}_{-3.18}$ & $      8.34^{+0.86}_{-0.86}$ &   2.83 &$  198.00^{+1.32}_{-1.32}$ & $  109.00^{+1.46}_{-1.46}$ & (1) \\
060418     & 1.489   & $    535.13^{+62.23}_{-77.16}$ & $     15.68^{+0.91}_{-0.96}$ & $      2.54^{+0.41}_{-0.42}$ &   2.18 &$  228.00^{+2.01}_{-2.09}$ & $   96.00^{+1.43}_{-1.50}$ & (3) \\
060605     & 3.78    & $    490.00^{+251.00}_{-251.00}$$^\star$ & $      4.61^{+0.60}_{-0.60}$ & $      1.09^{+0.29}_{-0.29}$ &   2.60 &$  174.00^{+3.52}_{-3.52}$ & $   65.00^{+2.18}_{-2.18}$ & (3) \\
060607A    & 3.0749  & $    575.00^{+200.00}_{-200.00}$$^\star$ & $     13.47^{+0.59}_{-0.59}$ & $      2.40^{+0.22}_{-0.22}$ &   2.24 &$  255.00^{+1.97}_{-1.97}$ & $  101.00^{+1.17}_{-1.17}$ & (3) \\
060904B    & 0.703   & $    135.00^{+41.00}_{-41.00}$$^\star$ & $      0.50^{+0.04}_{-0.04}$ & $      0.09^{+0.01}_{-0.01}$ &   2.67 &$   84.00^{+4.04}_{-4.04}$ & $   28.00^{+1.06}_{-1.06}$ & (3) \\
061007     & 1.261   & $    979.01^{+20.35}_{-20.35}$ & $     94.19^{+3.45}_{-3.45}$ & $     16.63^{+1.51}_{-1.51}$ &   1.89 &$  352.00^{+1.75}_{-1.75}$ & $  173.00^{+1.60}_{-1.60}$ & (3) \\
061121     & 1.314   & $   1404.60^{+106.44}_{-120.33}$ & $     25.61^{+0.96}_{-0.96}$ & $     22.32^{+1.64}_{-1.64}$ &   2.21 &$  230.00^{+1.07}_{-1.07}$ & $  105.00^{+0.98}_{-0.98}$ & (1) \\
070110     & 2.352   & $    370.00^{+170.00}_{-170.00}$$^\star$ & $      5.13^{+0.34}_{-0.34}$ & $      0.49^{+0.10}_{-0.10}$ &   3.07 &$  103.00^{+0.86}_{-0.86}$ & $   47.00^{+0.78}_{-0.78}$ & (1) \\
070318     & 0.836   & $    283.26^{+4.16}_{-4.16}$$^\diamond$ & $      1.31^{+0.06}_{-0.06}$ & $      0.14^{+0.01}_{-0.01}$ &   2.47 &$  116.00^{+4.13}_{-4.13}$ & $   40.00^{+1.06}_{-1.06}$ & (3) \\
070411     & 2.954   & $    328.04^{+8.45}_{-8.45}$$^\diamond$ & $     11.70^{+0.67}_{-0.67}$ & $      1.12^{+0.16}_{-0.16}$ &   2.65 &$  174.00^{+1.44}_{-1.44}$ & $   76.00^{+1.12}_{-1.12}$ & (3) \\
070419A    & 0.97    & $     27.00^{+16.00}_{-19.00}$$^\star$ & $      0.38^{+0.06}_{-0.06}$ & $      0.01^{+0.01}_{-0.01}$ &   2.77 &$   79.00^{+1.80}_{-1.80}$ & $   25.00^{+0.97}_{-0.97}$ & (3) \\
070802     & 2.45    & $    302.84^{+0.00}_{-0.00}$$^\ast$ & $      0.90^{+0.16}_{-0.16}$ & $      0.32^{+0.08}_{-0.08}$ &   3.34 &$   66.00^{+1.47}_{-1.47}$ & $   26.00^{+1.17}_{-1.17}$ & (4) \\
071010A    & 0.98    & $     61.00^{+0.00}_{-0.00}$$^\ast$ & $      0.12^{+0.02}_{-0.02}$ & $      0.05^{+0.02}_{-0.02}$ &   2.57 &$   81.00^{+2.86}_{-2.86}$ & $   21.00^{+1.12}_{-1.12}$ & (3) \\
071010B    & 0.947   & $    110.98^{+11.68}_{-9.73}$ & $      1.33^{+0.09}_{-0.09}$ & $      0.69^{+0.12}_{-0.12}$ &   2.46 &$  120.00^{+22.76}_{-22.76}$ & $   42.00^{+5.29}_{-5.29}$ & (2) \\
071025     & 5.2     & $    436.11^{+15.87}_{-15.87}$$^\diamond$ & $     64.71^{+1.99}_{-1.99}$ & $      6.86^{+0.86}_{-0.86}$ &   2.74 &$  236.00^{+4.78}_{-4.78}$ & $  125.00^{+1.91}_{-1.91}$ & (2) \\
071031     & 2.692   & $    150.59^{+0.00}_{-0.00}$$^\ast$ & $      3.24^{+0.47}_{-0.47}$ & $      0.41^{+0.08}_{-0.08}$ &   3.08 &$   99.00^{+1.80}_{-1.80}$ & $   42.00^{+1.53}_{-1.53}$ & (2) \\
071112C    & 0.823   & $    422.00^{+137.00}_{-87.00}$$^\star$ & $      1.88^{+0.25}_{-0.25}$ & $      0.78^{+0.10}_{-0.10}$ &   2.25 &$  146.00^{+4.69}_{-4.69}$ & $   50.00^{+1.91}_{-1.91}$ & (2) \\
080310     & 2.42    & $     94.32^{+6.47}_{-6.47}$$^\diamond$ & $      7.22^{+0.63}_{-0.63}$ & $      0.81^{+0.13}_{-0.13}$ &   2.26 &$  216.00^{+5.79}_{-5.79}$ & $   82.00^{+2.22}_{-2.22}$ & (2) \\
080319B    & 0.937   & $   1284.23^{+15.50}_{-15.50}$ & $    132.35^{+1.63}_{-1.63}$ & $     10.58^{+0.61}_{-0.61}$ &   1.24 &$  615.00^{+0.94}_{-0.94}$ & $  266.00^{+0.82}_{-0.82}$ & (1) \\
080319C    & 1.95    & $   1864.40^{+333.35}_{-472.00}$ & $     12.63^{+1.47}_{-1.47}$ & $     11.08^{+1.99}_{-1.99}$ &   2.54 &$  173.00^{+3.08}_{-3.08}$ & $   77.00^{+2.30}_{-2.30}$ & (3) \\
080330     & 1.51    & $     54.81^{+0.00}_{-0.00}$$^\diamond$ & $      0.48^{+0.11}_{-0.11}$ & $      0.18^{+0.04}_{-0.04}$ &   2.79 &$   87.00^{+2.71}_{-2.71}$ & $   28.00^{+1.68}_{-1.68}$ & (3) \\
080710     & 0.845   & $    300.00^{+500.00}_{-200.00}$$^\star$ & $      0.77^{+0.11}_{-0.11}$ & $      0.08^{+0.02}_{-0.02}$ &   3.29 &$   54.00^{+0.96}_{-0.96}$ & $   22.00^{+0.79}_{-0.79}$ & (2) \\
080804     & 2.2     & $    810.00^{+45.00}_{-45.00}$$^\diamond$ & $     14.49^{+0.80}_{-0.80}$ & $      3.42^{+0.44}_{-0.44}$ &   1.80 &$  344.00^{+2.39}_{-2.39}$ & $  125.00^{+1.73}_{-1.73}$ & (1) \\
080810     & 3.35    & $   2240.25^{+478.50}_{-726.45}$ & $     56.16^{+6.23}_{-6.57}$ & $     13.68^{+2.59}_{-2.64}$ &   2.07 &$  363.00^{+5.55}_{-5.79}$ & $  162.00^{+4.55}_{-4.79}$ & (3) \\
080916C(L) & 4.35    & $   3070.90^{+203.30}_{-230.05}$ & $    350.22^{+27.15}_{-27.15}$ & $    193.18^{+26.14}_{-26.14}$ &   0.79 &$ 1486.00^{+14.40}_{-14.40}$ & $  563.00^{+10.91}_{-10.91}$ & (1) \\
080928     & 1.692   & $    197.19^{+5.49}_{-5.49}$$^\diamond$ & $      4.12^{+0.33}_{-0.33}$ & $      0.65^{+0.03}_{-0.03}$ &   3.36 &$   72.00^{+0.72}_{-0.72}$ & $   36.00^{+0.71}_{-0.71}$ & (1) \\
081007A    & 0.5295  & $     93.30^{+22.94}_{-22.94}$ & $      0.15^{+0.01}_{-0.01}$ & $      0.06^{+0.01}_{-0.01}$ &   2.09 &$  115.00^{+1.20}_{-1.20}$ & $   28.00^{+0.59}_{-0.59}$ & (1) \\
081008     & 1.9685  & $    267.00^{+335.00}_{-62.00}$$^\star$ & $      9.59^{+0.45}_{-0.45}$ & $      0.63^{+0.05}_{-0.05}$ &   2.21 &$  223.00^{+1.65}_{-1.65}$ & $   87.00^{+1.05}_{-1.05}$ & (2) \\
081109A    & 0.9787  & $    474.89^{+118.72}_{-118.72}$ & $      2.27^{+0.15}_{-0.15}$ & $      0.25^{+0.02}_{-0.02}$ &   2.75 &$  101.00^{+8.61}_{-8.61}$ & $   40.00^{+2.39}_{-2.39}$ & (2) \\
081203A    & 2.05    & $   1541.00^{+757.00}_{-757.00}$$^\star$ & $     41.84^{+1.63}_{-1.63}$ & $      4.31^{+0.30}_{-0.30}$ &   2.56 &$  200.00^{+1.01}_{-1.01}$ & $  104.00^{+1.01}_{-1.01}$ & (3) \\
090102     & 1.547   & $   1100.30^{+91.69}_{-106.97}$ & $     16.72^{+0.87}_{-0.87}$ & $      8.74^{+1.08}_{-1.08}$ &   1.70 &$  351.00^{+3.48}_{-3.48}$ & $  130.00^{+1.80}_{-1.80}$ & (2) \\
090313     & 3.38    & $    341.89^{+0.00}_{-0.00}$$^\ast$ & $      7.44^{+1.06}_{-1.06}$ & $      1.32^{+0.50}_{-0.50}$ &   3.03 &$  123.00^{+2.20}_{-2.20}$ & $   56.00^{+2.01}_{-2.01}$ & (1) \\
090323(L)  & 3.57    & $   2073.16^{+107.84}_{-107.84}$ & $    442.17^{+4.94}_{-4.94}$ & $     42.04^{+3.17}_{-3.17}$ &   2.30 &$  392.00^{+0.55}_{-0.55}$ & $  241.00^{+0.67}_{-0.67}$ & (1) \\
090418A    & 1.608   & $   1220.54^{+192.99}_{-268.62}$ & $     15.64^{+1.26}_{-1.31}$ & $      3.10^{+0.64}_{-0.64}$ &   2.20 &$  227.00^{+2.28}_{-2.37}$ & $   96.00^{+1.93}_{-2.00}$ & (4) \\
090510 & 0.903   & $   8995.59^{+664.11}_{-664.11}$ & $      3.91^{+0.09}_{-0.09}$ & $     26.58^{+1.03}_{-1.03}$ &   3.20 &$   72.00^{+11.10}_{-11.10}$ & $   35.00^{+3.64}_{-3.64}$ & (2) \\
090618     & 0.54    & $    229.52^{+5.06}_{-5.06}$ & $     28.24^{+0.16}_{-0.16}$ & $      1.39^{+0.09}_{-0.09}$ &   1.96 &$  247.00^{+0.17}_{-0.17}$ & $  112.00^{+0.15}_{-0.15}$ & (1) \\
090726     & 2.71    & $    114.29^{+6.71}_{-6.71}$$^\diamond$ & $      3.21^{+0.37}_{-0.37}$ & $      0.58^{+0.16}_{-0.16}$ &   2.73 &$  135.00^{+1.96}_{-1.96}$ & $   52.00^{+1.51}_{-1.51}$ & (5) \\
090812     & 2.452   & $   1325.57^{+362.46}_{-907.88}$ & $     21.91^{+0.66}_{-0.66}$ & $     12.52^{+3.31}_{-3.31}$ &   1.68 &$  413.00^{+1.56}_{-1.56}$ & $  151.00^{+1.14}_{-1.14}$ & (1) \\
090902B(L) & 1.822   & $   2978.45^{+45.90}_{-45.90}$ & $    355.48^{+1.03}_{-1.03}$ & $     99.20^{+1.90}_{-1.90}$ &   0.93 &$ 1034.00^{+0.37}_{-0.37}$ & $  443.00^{+0.32}_{-0.32}$ & (1) \\
090926A(L) & 2.1062  & $   1036.90^{+18.14}_{-18.14}$ & $    214.71^{+1.00}_{-1.00}$ & $    112.89^{+1.67}_{-1.67}$ &   0.91 &$ 1025.00^{+0.60}_{-0.60}$ & $  405.00^{+0.47}_{-0.47}$ & (1) \\
091020     & 1.71    & $    660.80^{+74.44}_{-74.44}$ & $      7.97^{+0.40}_{-0.40}$ & $      3.54^{+0.44}_{-0.44}$ &   2.13 &$  226.00^{+1.42}_{-1.42}$ & $   85.00^{+1.07}_{-1.07}$ & (1) \\
091024     & 1.092   & $    794.00^{+231.00}_{-231.00}$$^\star$ & $      7.39^{+0.36}_{-0.36}$ & $      0.44^{+0.07}_{-0.07}$ &   2.65 &$  129.00^{+0.80}_{-0.80}$ & $   58.00^{+0.72}_{-0.72}$ & (6) \\
091029     & 2.752   & $    230.37^{+65.66}_{-65.66}$ & $      8.66^{+0.36}_{-0.36}$ & $      2.04^{+0.11}_{-0.11}$ &   2.61 &$  170.00^{+0.89}_{-0.89}$ & $   72.00^{+0.75}_{-0.75}$ & (1) \\
100414A(L) & 1.368   & $   1582.14^{+34.63}_{-34.63}$ & $     73.22^{+0.52}_{-0.52}$ & $     17.02^{+0.63}_{-0.63}$ &   1.54 &$  469.00^{+0.42}_{-0.42}$ & $  201.00^{+0.36}_{-0.36}$ & (1) \\
100621A    & 0.542   & $    163.45^{+12.34}_{-13.88}$ & $      3.75^{+0.16}_{-0.16}$ & $      0.41^{+0.08}_{-0.08}$ &   3.73 &$   42.00^{+0.22}_{-0.22}$ & $   24.00^{+0.26}_{-0.26}$ & (7) \\
100728B    & 2.106   & $    476.30^{+57.33}_{-57.33}$ & $      3.68^{+0.25}_{-0.25}$ & $      4.38^{+0.66}_{-0.66}$ &   1.53 &$  361.00^{+3.12}_{-3.12}$ & $  103.00^{+1.77}_{-1.77}$ & (1) \\
100814A    & 1.44    & $    357.89^{+16.47}_{-16.47}$ & $      8.64^{+0.29}_{-0.29}$ & $      1.82^{+0.37}_{-0.37}$ &   2.77 &$  126.00^{+3.26}_{-3.26}$ & $   59.00^{+1.11}_{-1.11}$ & (1) \\
100901A    & 1.408   & $    305.63^{+4.25}_{-4.25}$$^\diamond$ & $      2.81^{+0.40}_{-0.40}$ & $      0.20^{+0.05}_{-0.05}$ &   3.10 &$   82.00^{+2.36}_{-2.36}$ & $   36.00^{+1.41}_{-1.41}$ & (2) \\
100906A    & 1.727   & $    204.29^{+66.17}_{-66.17}$ & $     30.02^{+0.79}_{-0.79}$ & $      7.19^{+0.69}_{-0.69}$ &   2.00 &$  298.00^{+4.53}_{-4.53}$ & $  128.00^{+1.52}_{-1.52}$ & (2) \\
110205A    & 2.22    & $    715.00^{+239.00}_{-239.00}$$^\star$ & $     64.77^{+0.00}_{-0.00}$ & $      3.73^{+0.21}_{-0.21}$ &   2.91 &$  159.00^{+0.00}_{-0.00}$ & $   96.00^{+0.00}_{-0.00}$ & (1) \\
110213A    & 1.46    & $    276.92^{+29.08}_{-29.08}$ & $      8.30^{+0.34}_{-0.34}$ & $      3.59^{+0.38}_{-0.38}$ &   2.51 &$  158.00^{+0.82}_{-0.82}$ & $   68.00^{+0.70}_{-0.70}$ & (1) \\
110731A(L) & 2.83    & $   1234.07^{+64.79}_{-64.79}$ & $     50.29^{+0.81}_{-0.81}$ & $     45.93^{+3.33}_{-3.33}$ &   0.70 &$ 1113.00^{+2.25}_{-2.25}$ & $  336.00^{+1.36}_{-1.36}$ & (1) \\
110801A    & 1.858   & $    208.63^{+34.30}_{-71.45}$ & $      9.05^{+1.13}_{-1.67}$ & $      1.93^{+0.66}_{-0.74}$ &   2.64 &$  150.00^{+2.34}_{-3.46}$ & $   66.00^{+2.07}_{-3.06}$ & (5) \\
120119A    & 1.728   & $    529.23^{+27.28}_{-30.01}$ & $     35.46^{+2.55}_{-2.55}$ & $     10.99^{+1.61}_{-1.61}$ &   3.00 &$  129.00^{+1.16}_{-1.16}$ & $   75.00^{+1.36}_{-1.36}$ & (7) \\
120404A    & 2.876   & $    242.04^{+7.74}_{-7.74}$$^\diamond$ & $      6.44^{+0.40}_{-0.40}$ & $      1.27^{+0.21}_{-0.21}$ &   3.40 &$   84.00^{+0.66}_{-0.66}$ & $   43.00^{+0.67}_{-0.67}$ & (7) \\
120711A    & 1.405   & $   3168.63^{+101.63}_{-101.63}$ & $    188.11^{+0.96}_{-0.96}$ & $     28.98^{+1.82}_{-1.82}$ &   2.38 &$  259.00^{+8.16}_{-8.16}$ & $  158.00^{+3.33}_{-3.33}$ & (1) \\
120815A    & 2.358   & $     97.07^{+5.78}_{-5.78}$$^\diamond$ & $      1.46^{+0.21}_{-0.21}$ & $      1.29^{+0.18}_{-0.18}$ &   2.64 &$  127.00^{+2.27}_{-2.27}$ & $   44.00^{+1.57}_{-1.57}$ & (1) \\
120909A    & 3.93    & $    984.27^{+119.70}_{-119.70}$ & $     72.16^{+2.28}_{-2.28}$ & $     14.26^{+3.29}_{-3.29}$ &   2.46 &$  280.00^{+1.11}_{-1.11}$ & $  142.00^{+1.12}_{-1.12}$ & (1) \\
120922A    & 3.1     & $    247.02^{+24.60}_{-24.60}$ & $     19.10^{+1.08}_{-1.08}$ & $      5.21^{+1.32}_{-1.32}$ &   2.95 &$  145.00^{+1.02}_{-1.02}$ & $   73.00^{+1.03}_{-1.03}$ & (1) \\
121011A    & (2.0)     & $   1129.70^{+413.06}_{-413.06}$ & $     13.88^{+0.56}_{-0.56}$ & $      5.31^{+1.47}_{-1.47}$ &   2.76 &$  146.00^{+4.85}_{-4.85}$ & $   70.00^{+1.69}_{-1.69}$ & (8) \\
121128A    & 2.2     & $    192.25^{+12.32}_{-12.32}$ & $     13.83^{+0.48}_{-0.48}$ & $      9.38^{+0.84}_{-0.84}$ &   1.87 &$  322.00^{+1.40}_{-1.40}$ & $  119.00^{+1.03}_{-1.03}$ & (1) \\
130215A    & 0.597   & $    335.29^{+67.57}_{-67.57}$ & $      3.83^{+0.09}_{-0.09}$ & $      0.31^{+0.07}_{-0.07}$ &   2.87 &$   89.00^{+0.27}_{-0.27}$ & $   41.00^{+0.25}_{-0.25}$ & (1) \\
130420A    & 1.297   & $    130.88^{+6.95}_{-6.95}$ & $      3.37^{+0.12}_{-0.12}$ & $      0.98^{+0.22}_{-0.22}$ &   2.55 &$  132.00^{+0.59}_{-0.59}$ & $   52.00^{+0.46}_{-0.46}$ & (1) \\
130427A(L) & 0.3399  & $   1432.35^{+12.06}_{-12.06}$ & $     77.52^{+0.41}_{-0.41}$ & $     24.71^{+0.58}_{-0.58}$ &   1.34 &$  455.00^{+0.30}_{-0.30}$ & $  199.00^{+0.26}_{-0.26}$ & (1) \\
130610A    & 2.092   & $    881.19^{+190.50}_{-190.50}$ & $      8.73^{+0.72}_{-0.72}$ & $      2.54^{+0.75}_{-0.75}$ &   2.31 &$  205.00^{+2.12}_{-2.12}$ & $   81.00^{+1.68}_{-1.68}$ & (1) \\
130612A    & 2.006   & $    173.53^{+30.06}_{-30.06}$ & $      0.64^{+0.06}_{-0.06}$ & $      1.37^{+0.36}_{-0.36}$ &   2.04 &$  185.00^{+2.32}_{-2.32}$ & $   49.00^{+1.23}_{-1.23}$ & (1) \\
130831A    & 0.4791  & $     79.87^{+13.31}_{-10.35}$ & $      0.65^{+0.03}_{-0.03}$ & $      0.25^{+0.03}_{-0.03}$ &   2.86 &$   70.00^{+0.39}_{-0.39}$ & $   26.00^{+0.29}_{-0.29}$ & (1) \\
131231A    & 0.642   & $    292.42^{+6.62}_{-6.62}$ & $     22.29^{+0.10}_{-0.10}$ & $      2.26^{+0.07}_{-0.07}$ &   2.00 &$  238.00^{+0.13}_{-0.13}$ & $  105.00^{+0.11}_{-0.11}$ & (1) \\
140423A    & 3.26    & $    494.90^{+67.68}_{-67.68}$ & $     68.68^{+1.93}_{-1.93}$ & $     10.71^{+2.11}_{-2.11}$ &   2.30 &$  302.00^{+1.06}_{-1.06}$ & $  148.00^{+1.04}_{-1.04}$ & (1) \\
140512A    & 0.725   & $   1193.45^{+98.41}_{-98.41}$ & $     10.16^{+0.16}_{-0.16}$ & $      1.08^{+0.09}_{-0.09}$ &   2.37 &$  160.00^{+0.32}_{-0.32}$ & $   71.00^{+0.28}_{-0.28}$ & (9) \\
140629A    & 2.275   & $    200.57^{+6.83}_{-6.83}$$^\diamond$ & $      6.53^{+0.54}_{-0.54}$ & $      2.53^{+0.24}_{-0.24}$ &   2.18 &$  226.00^{+2.36}_{-2.36}$ & $   83.00^{+1.72}_{-1.72}$ & (1) \\
141109A    & 2.993   & $    650.86^{+83.85}_{-111.80}$ & $     40.63^{+3.32}_{-3.90}$ & $     10.98^{+1.85}_{-1.90}$ &   2.98 &$  154.00^{+1.57}_{-1.84}$ & $   87.00^{+1.77}_{-2.07}$ & (1) \\
141221A    & 1.452   & $    446.44^{+78.39}_{-78.39}$ & $      2.08^{+0.19}_{-0.19}$ & $      1.90^{+0.32}_{-0.32}$ &   2.04 &$  198.00^{+2.28}_{-2.28}$ & $   63.00^{+1.44}_{-1.44}$ & (1) \\
150910A    & 1.359   & $    382.80^{+4.81}_{-4.81}$$^\diamond$ & $      6.61^{+0.55}_{-0.55}$ & $      0.29^{+0.11}_{-0.11}$ &   3.16 &$   86.00^{+1.44}_{-1.44}$ & $   43.00^{+0.98}_{-0.98}$ & (10) \\
160629A    & 3.332   & $   1259.52^{+83.48}_{-83.48}$ & $     56.03^{+1.81}_{-1.81}$ & $     15.32^{+2.04}_{-2.04}$ &   1.91 &$  415.00^{+1.68}_{-1.68}$ & $  177.00^{+1.43}_{-1.43}$ & (1) \\
161023A    & 2.708   & $    604.40^{+137.20}_{-103.82}$ & $     68.17^{+9.74}_{-9.74}$ & $     31.90^{+4.81}_{-4.81}$ &   2.19 &$  314.00^{+5.81}_{-5.81}$ & $  152.00^{+5.45}_{-5.45}$ & (11) \\
180325A    & 2.248   & $    993.89^{+162.40}_{-126.67}$ & $     21.83^{+3.45}_{-3.45}$ & $     30.60^{+4.85}_{-4.85}$ &   2.00 &$  306.00^{+6.05}_{-6.05}$ & $  124.00^{+4.88}_{-4.88}$ & (12) \\
180720B(L) & 0.654   & $   1070.76^{+15.63}_{-15.63}$ & $     58.33^{+0.37}_{-0.37}$ & $      9.76^{+0.15}_{-0.15}$ &   1.89 &$  295.00^{+11.36}_{-11.36}$ & $  142.00^{+3.65}_{-3.65}$ & (13) \\
191016A    & (2.0)     & $    354.83^{+6.59}_{-6.59}$$^\diamond$ & $     15.93^{+0.98}_{-0.98}$ & $      0.85^{+0.11}_{-0.11}$ &   3.16 &$  105.00^{+0.81}_{-0.81}$ & $   57.00^{+0.88}_{-0.88}$ & (14) \\
201015A    & 0.426   & $     15.00^{+1.85}_{-1.85}$$^\diamond$ & $      0.03^{+0.01}_{-0.01}$ & $      0.02^{+0.00}_{-0.00}$ &   2.27 &$   77.00^{+3.94}_{-3.94}$ & $   16.00^{+1.27}_{-1.27}$ & (15) \\
201216     & 1.1     & $    739.85^{+26.75}_{-26.75}$ & $     54.01^{+0.03}_{-0.03}$ & $      8.48^{+0.34}_{-0.34}$ &   2.26 &$  234.00^{+24.37}_{-24.37}$ & $  120.00^{+8.35}_{-8.35}$ & (15) \\
201223A    & (2.0)     & $    337.55^{+17.92}_{-17.92}$ & $      2.11^{+0.41}_{-0.41}$ & $      2.08^{+0.49}_{-0.49}$ &   1.72 &$  284.00^{+61.79}_{-61.79}$ & $   80.00^{+12.11}_{-12.11}$ & (16) \\
\enddata
\end{deluxetable*}
\vspace{-40pt}
\begin{itemize}
    \item Notes.
    \item[$^a$] The measured redshifts are taken from the website \url{https://www.mpe.mpg.de/~jcg/grbgen.html}. When the redshift is unknown, we adopt the mean value of the redshift distribution of GRBs with known redshifts, i.e., $z=2$.
    \item[$^b$] The $E_{\rm p}$ values of bursts marked with “$\star$” and “$\ast$” are taken from \citet{2015ApJ...813..116L} and \citet{2025ApJ...Sun}, respectively. Data marked with “$\diamond$” are derived from the empirical relation in \citet{2007ApJ...655L..25Z}.
    \item[$^c$] The values of $t_{\rm p}$ obtained from the reference.
    \item Reference. (1) \citet{2018A&A...609A.112G}; (2) \citet{2013ApJ...774...13L}; (3) \citet{2010ApJ...725.2209L}; (4) \citet{2011MNRAS.414.3537P}; (5) \citet{2015MNRAS.454.3567Y}; (6) \citet{2014MNRAS.445.1625N}; (7) \citet{2017JHEAp..13....1Y}; (8) \citet{2016RAA....16...12X}; (9) \citet{2016ApJ...833..100H}; (10) \citet{2020ApJ...896....4X}; (11) \citet{2018A&A...620A.119D}; (12) \citet{2021ApJ...908...39B}; (13) \citet{2020A&A...636A..55R}; (14) \citet{2022MNRAS.511.6205P}; (15) \citet{2023ApJ...942...34R}; (16) \citet{2023NatAs...7..724X}.

\end{itemize}
\end{longrotatetable}

\clearpage

\begin{longrotatetable}
\begin{deluxetable*}{lcccccccccc}
\tablecaption{The Parameters of the GRBs with Observed Jet Break features. \label{tab:tab2}}
\tablewidth{750pt}
\tabletypesize{\small}
\tablehead{
\multicolumn{1}{l}{GRBs} & \colhead{$T^{a}_{\rm jet}$} & \colhead{$\theta$[ISM]} & \colhead{$\theta$[Wind]} & \colhead{$E^{b}_{\rm \gamma,52}$[ISM]}& \colhead{$E^{b}_{\rm \gamma,52}$[Wind]} & \colhead{$L^{c}_{\rm \gamma,52}$[ISM]} & \colhead{$L^{c}_{\rm \gamma,52}$[Wind]} & \colhead{$a_{\rm j}$[ISM]}  & \colhead{$a_{\rm j}$[Wind]} & \colhead{Ref.} \\
\colhead{} & \colhead{(day)} & \colhead{(°)} & \colhead{(°)}  & \colhead{(erg)}& \colhead{(erg)} & \colhead{(erg/s)} & \colhead{(erg/s)} & \colhead{} & \colhead{} & \colhead{$T_{\rm jet}$} \\[-18pt]
}

\startdata
990123  & 2.04 $\pm$ 0.46 &$  3.61^{+0.31}_{-0.31}$ & $  2.12^{+0.12}_{-0.12}$ & 3.43E-01 & 1.19E-01 & 8.36E-02 & 2.89E-02 &  30.31 &    8.74 & (1) \\
050502A & 0.1 $\pm$ 0.01 &$  1.33^{+0.05}_{-0.05}$ & $  1.77^{+0.04}_{-0.04}$ & 2.55E-03 & 4.51E-03 & 1.68E-03 & 2.98E-03 &   9.12 &    3.93 & (1) \\
050730  & 0.12 $\pm$ 0.05 &$  1.30^{+0.20}_{-0.20}$ & $  1.56^{+0.16}_{-0.16}$ & 4.66E-03 & 6.74E-03 & 4.22E-04 & 6.11E-04 &   4.08 &    2.29 & (1) \\
050820A & 7.52 $\pm$ 2.31 &$  6.91^{+0.80}_{-0.80}$ & $  4.76^{+0.37}_{-0.37}$ & 1.31E-01 & 6.24E-02 & 2.88E-02 & 1.37E-02 &  22.54 &    7.15 & (1) \\
050904  & 2.6 $\pm$ 1.0 &$  2.95^{+0.43}_{-0.43}$ & $  2.10^{+0.20}_{-0.20}$ & 1.08E-01 & 5.50E-02 & 8.48E-03 & 4.30E-03 &  15.14 &    5.50 & (1) \\
050922C & 0.09 $\pm$ 0.001 &$  1.52^{+0.05}_{-0.05}$ & $  1.98^{+0.01}_{-0.01}$ & 2.89E-03 & 4.90E-03 & 6.84E-03 & 1.16E-02 &   6.43 &    3.10 & (1) \\
060124  & 0.68 $\pm$ 0.32 &$  2.76^{+0.49}_{-0.49}$ & $  2.41^{+0.28}_{-0.28}$ & 3.16E-02 & 2.42E-02 & 1.40E-02 & 1.07E-02 &   7.65 &    3.49 & (1) \\
060210  & 0.3 $\pm$ 0.11 &$  1.58^{+0.22}_{-0.22}$ & $  1.46^{+0.13}_{-0.13}$ & 2.27E-02 & 1.94E-02 & 3.19E-03 & 2.72E-03 &   5.47 &    2.78 & (2) \\
060418  & 0.07 $\pm$ 0.04 &$  1.40^{+0.30}_{-0.30}$ & $  1.68^{+0.24}_{-0.24}$ & 4.67E-03 & 6.76E-03 & 7.57E-04 & 1.09E-03 &   5.57 &    2.82 & (1) \\
060605  & 0.24 $\pm$ 0.02 &$  2.03^{+0.07}_{-0.07}$ & $  2.64^{+0.06}_{-0.06}$ & 2.88E-03 & 4.90E-03 & 6.84E-04 & 1.16E-03 &   6.15 &    3.00 & (2) \\
061121  & 0.26 $\pm$ 0.04 &$  2.21^{+0.13}_{-0.13}$ & $  2.10^{+0.08}_{-0.08}$ & 1.91E-02 & 1.73E-02 & 1.66E-02 & 1.51E-02 &   8.88 &    3.86 & (3) \\
070318  & 3.57 $\pm$ 0.63 &$  9.34^{+0.62}_{-0.62}$ & $  9.02^{+0.40}_{-0.40}$ & 1.74E-02 & 1.62E-02 & 1.83E-03 & 1.71E-03 &  18.91 &    6.30 & (1) \\
070411  & 0.24 $\pm$ 0.11 &$  1.94^{+0.33}_{-0.33}$ & $  2.19^{+0.25}_{-0.25}$ & 6.68E-03 & 8.58E-03 & 6.41E-04 & 8.24E-04 &   5.88 &    2.91 & (1) \\
070419A & 0.02 $\pm$ 0.0 &$  1.52^{+0.03}_{-0.03}$ & $  3.31^{+0.01}_{-0.01}$ & 1.33E-04 & 6.30E-04 & 5.23E-06 & 2.48E-05 &   2.10 &    1.45 & (1) \\
071010A & 0.81 $\pm$ 0.2 &$  7.05^{+0.68}_{-0.68}$ & $ 11.19^{+0.69}_{-0.69}$ & 8.81E-04 & 2.22E-03 & 4.14E-04 & 1.04E-03 &   9.96 &    4.10 & (1) \\
071010B & 3.44 $\pm$ 0.39 &$  9.00^{+0.39}_{-0.39}$ & $  8.79^{+0.25}_{-0.25}$ & 1.63E-02 & 1.56E-02 & 8.55E-03 & 8.15E-03 &  18.85 &    6.44 & (1) \\
080310  & 0.34 $\pm$ 0.04 &$  2.47^{+0.11}_{-0.11}$ & $  2.80^{+0.08}_{-0.08}$ & 6.74E-03 & 8.63E-03 & 7.60E-04 & 9.73E-04 &   9.33 &    4.01 & (1) \\
080319B & 0.03 $\pm$ 0.01 &$  0.86^{+0.11}_{-0.11}$ & $  0.85^{+0.07}_{-0.07}$ & 1.48E-02 & 1.46E-02 & 1.18E-03 & 1.17E-03 &   9.20 &    3.95 & (1) \\
080330  & 1.0 $\pm$ 0.0 &$  5.84^{+0.17}_{-0.17}$ & $  7.80^{+0.05}_{-0.05}$ & 2.50E-03 & 4.45E-03 & 9.45E-04 & 1.68E-03 &   8.87 &    3.81 & (1) \\
080710  & 0.23 $\pm$ 0.02 &$  3.56^{+0.13}_{-0.13}$ & $  5.18^{+0.11}_{-0.11}$ & 1.49E-03 & 3.16E-03 & 1.60E-04 & 3.39E-04 &   3.36 &    1.99 & (1) \\
080810  & 0.11 $\pm$ 0.02 &$  1.15^{+0.08}_{-0.08}$ & $  1.19^{+0.05}_{-0.05}$ & 1.12E-02 & 1.21E-02 & 2.74E-03 & 2.96E-03 &   7.26 &    3.37 & (1) \\
080928  & 0.14 $\pm$ 0.04 &$  2.08^{+0.22}_{-0.22}$ & $  2.74^{+0.20}_{-0.20}$ & 2.72E-03 & 4.71E-03 & 4.32E-04 & 7.48E-04 &   2.62 &    1.72 & (1) \\
081007A & 11.57 $\pm$ 0.0 &$ 20.33^{+0.21}_{-0.21}$ & $ 21.66^{+0.05}_{-0.05}$ & 9.56E-03 & 1.08E-02 & 3.54E-03 & 4.01E-03 &  40.80 &   10.59 & (1) \\
081008  & 0.21 $\pm$ 0.07 &$  2.10^{+0.26}_{-0.26}$ & $  2.40^{+0.20}_{-0.20}$ & 6.45E-03 & 8.38E-03 & 4.24E-04 & 5.51E-04 &   8.18 &    3.64 & (1) \\
081203A & 0.12 $\pm$ 0.02 &$  1.40^{+0.09}_{-0.09}$ & $  1.43^{+0.06}_{-0.06}$ & 1.26E-02 & 1.31E-02 & 1.29E-03 & 1.34E-03 &   4.90 &    2.60 & (1) \\
090313  & 0.94 $\pm$ 0.0 &$  3.29^{+0.06}_{-0.06}$ & $  3.37^{+0.01}_{-0.01}$ & 1.23E-02 & 1.29E-02 & 2.18E-03 & 2.29E-03 &   7.06 &    3.29 & (3) \\
090323(L) & 17.8 $\pm$ 1.7 &$  5.86^{+0.21}_{-0.21}$ & $  2.51^{+0.06}_{-0.06}$ & 2.31E+00 & 4.23E-01 & 2.19E-01 & 4.02E-02 &  40.06 &   10.54 & (1) \\
090618  & 0.08 $\pm$ 0.01 &$  1.64^{+0.08}_{-0.08}$ & $  1.69^{+0.05}_{-0.05}$ & 1.15E-02 & 1.23E-02 & 5.66E-04 & 6.06E-04 &   7.05 &    3.31 & (1) \\
090902B(L) & 6.2 $\pm$ 2.4 &$  4.86^{+0.70}_{-0.70}$ & $  2.29^{+0.22}_{-0.22}$ & 1.28E+00 & 2.85E-01 & 3.56E-01 & 7.94E-02 &  87.61 &   17.73 & (1) \\
090926A(L) & 4.06 $\pm$ 0.81 &$  4.26^{+0.32}_{-0.32}$ & $  2.28^{+0.11}_{-0.11}$ & 5.92E-01 & 1.71E-01 & 3.11E-01 & 8.97E-02 &  76.13 &   16.14 & (1) \\
091029  & 0.03 $\pm$ 0.0 &$  0.94^{+0.00}_{-0.00}$ & $  1.43^{+0.00}_{-0.00}$ & 1.17E-03 & 2.68E-03 & 2.74E-04 & 6.31E-04 &   2.79 &    1.79 & (1) \\
100814A & 2.0 $\pm$ 0.07 &$  5.34^{+0.07}_{-0.07}$ & $  4.54^{+0.04}_{-0.04}$ & 3.75E-02 & 2.71E-02 & 7.90E-03 & 5.71E-03 &  11.74 &    4.67 & (1) \\
100901A & 11.57 $\pm$ 0.0 &$ 11.93^{+0.21}_{-0.21}$ & $  9.35^{+0.03}_{-0.03}$ & 6.06E-02 & 3.73E-02 & 4.35E-03 & 2.68E-03 &  17.07 &    5.88 & (1) \\
100906A & 0.59 $\pm$ 0.05 &$  2.77^{+0.09}_{-0.09}$ & $  2.38^{+0.05}_{-0.05}$ & 3.51E-02 & 2.59E-02 & 8.42E-03 & 6.22E-03 &  14.42 &    5.32 & (1) \\
110205A & 1.2 $\pm$ 0.0 &$  3.09^{+0.00}_{-0.00}$ & $  2.25^{+0.00}_{-0.00}$ & 9.40E-02 & 5.00E-02 & 5.42E-03 & 2.88E-03 &   8.57 &    3.77 & (1) \\
110801A & 0.18 $\pm$ 0.1 &$  2.03^{+0.42}_{-0.42}$ & $  2.36^{+0.33}_{-0.33}$ & 5.67E-03 & 7.69E-03 & 1.21E-03 & 1.64E-03 &   5.31 &    2.72 & (1) \\
120119A & 0.13 $\pm$ 0.02 &$  1.54^{+0.09}_{-0.09}$ & $  1.57^{+0.06}_{-0.06}$ & 1.28E-02 & 1.32E-02 & 3.97E-03 & 4.10E-03 &   3.47 &    2.05 & (1) \\
120404A & 0.06 $\pm$ 0.0 &$  1.25^{+0.01}_{-0.01}$ & $  1.81^{+0.00}_{-0.00}$ & 1.53E-03 & 3.21E-03 & 3.02E-04 & 6.33E-04 &   1.83 &    1.36 & (1) \\
130427A(L) & 0.43 $\pm$ 0.05 &$  2.85^{+0.12}_{-0.12}$ & $  2.07^{+0.06}_{-0.06}$ & 9.62E-02 & 5.08E-02 & 3.07E-02 & 1.62E-02 &  22.67 &    7.20 & (1) \\
140512A & 0.21 $\pm$ 0.02 &$  2.56^{+0.09}_{-0.09}$ & $  2.71^{+0.06}_{-0.06}$ & 1.01E-02 & 1.13E-02 & 1.08E-03 & 1.20E-03 &   7.15 &    3.35 & (1) \\
140629A & 0.42 $\pm$ 0.09 &$  2.76^{+0.22}_{-0.22}$ & $  3.06^{+0.16}_{-0.16}$ & 7.56E-03 & 9.32E-03 & 2.92E-03 & 3.60E-03 &  10.87 &    4.43 & (1) \\
180720B(L) & 1.08 $\pm$ 0.27 &$  3.86^{+0.36}_{-0.36}$ & $  2.66^{+0.17}_{-0.17}$ & 1.32E-01 & 6.28E-02 & 2.22E-02 & 1.05E-02 &  19.88 &    6.59 & (1) \\
\enddata
\end{deluxetable*}
\vspace{-40pt}
\begin{itemize}
    \item Notes.
    \item[$^a$] The values of $T_{\rm jet}$ obtained from the Reference.
    \item[$^b$] $E_{\rm \gamma,52}$[ISM] and $E_{\rm \gamma,52}$[Wind] represent the gamma-ray energies corrected for the jet half-opening angle in the ISM and wind media, respectively.
    \item[$^c$] $L_{\rm \gamma,52}$[ISM] and $L_{\rm \gamma,52}$[Wind] represent the gamma-ray luminosity corrected by the jet half-opening angle in the ISM and Wind media, respectively.
    \item Reference. 
    (1) \citet{2020ApJ...900..112Z}; 
    (2) \citet{2012ApJ...745..168L}; 
    (3) \href{https://www.swift.ac.uk/xrt\_live\_cat/}{https://www.swift.ac.uk/xrt\_live\_cat/}.
\end{itemize}
\end{longrotatetable}

\clearpage
\renewcommand{\arraystretch}{1.2}
\begin{deluxetable*}{llccc}[h]
\centering
\tablecaption{The Results of the Regression Analysis for the Correlations in the GRBs Sample with Onset Bump Features, Where $r$ Is the Correlation Coefficient, $p$ Is the Chance Probability, and $\sigma_{\rm int}$ Is the Dispersion.
\label{tab:tab3}} 
\tablewidth{780pt}
\tabletypesize{\small}
\tablehead{
\multicolumn{1}{l}{Relations} & \multicolumn{1}{l}{Expressions} & \colhead{$r$} & \colhead{$p$} & \colhead{$\sigma_{\rm int}$}\\
[-18pt]
}
\startdata
\multicolumn{5}{l}{\textbf{Fitting results from 67 GRBs with well-measured spectra:}}\\
$\Gamma_0$[ISM]($E_{\rm iso,52}$) &  log$\Gamma_0$ = (2.00 $\pm$ 0.05) + (0.29 $\pm$ 0.03) $\times$ log$E_{\rm iso,52}$ & 0.74 & $<10^{-4}$ & 0.21 \\
$\Gamma_0$[Wind]($E_{\rm iso,52}$) &  log$\Gamma_0$ = (1.54 $\pm$ 0.03) + (0.36 $\pm$ 0.02) $\times$ log$E_{\rm iso,52}$ & 0.90 & $<10^{-4}$ & 0.14 \\
$\Gamma_0$[ISM]($L_{\rm iso,52}$) &  log$\Gamma_0$ = (2.17 $\pm$ 0.03) + (0.28 $\pm$ 0.03) $\times$ log$L_{\rm iso,52}$ & 0.78 & $<10^{-4}$ & 0.19 \\
$\Gamma_0$[Wind]($L_{\rm iso,52}$) &  log$\Gamma_0$ = (1.77 $\pm$ 0.02) + (0.32 $\pm$ 0.02) $\times$ log$L_{\rm iso,52}$ & 0.86 & $<10^{-4}$ & 0.16 \\
$\Gamma_0$[ISM]($E_{\rm p,z}$) &  log$\Gamma_0$ = (1.34 $\pm$ 0.20) + (0.36 $\pm$ 0.07) $\times$ log$E_{\rm p,z}$ & 0.53 & $<10^{-4}$ & 0.26 \\
$\Gamma_0$[Wind]($E_{\rm p,z}$) &  log$\Gamma_0$ = (0.76 $\pm$ 0.19) + (0.43 $\pm$ 0.07) $\times$ log$E_{\rm p,z}$ & 0.63 & $<10^{-4}$ & 0.24 \\
$E_{\rm iso,52}$($E_{\rm p,z}$, $\Gamma_0$[ISM]) &  log$E_{\rm iso,52}$ = (-3.81 $\pm$ 0.46) + (0.70 $\pm$ 0.15) $\times$ log$E_{\rm p,z}$ + (1.30 $\pm$ 0.22) $\times$ log$\Gamma_0$ & 0.81 & $<10^{-4}$  & 0.46 \\
$E_{\rm iso,52}$($E_{\rm p,z}$, $\Gamma_0$[Wind]) &  log$E_{\rm iso,52}$ = (-3.51 $\pm$ 0.29) + (0.34 $\pm$ 0.12) $\times$ log$E_{\rm p,z}$ + (1.90 $\pm$ 0.17) $\times$ log$\Gamma_0$ & 0.91 & $<10^{-4}$  & 0.33 \\
$L_{\rm iso,52}$($E_{\rm p,z}$, $\Gamma_0$[ISM])  &  log$L_{\rm iso,52}$ = (-5.19 $\pm$ 0.40) + (0.92 $\pm$ 0.13) $\times$ log$E_{\rm p,z}$ + (1.38 $\pm$ 0.19) $\times$ log$\Gamma_0$ & 0.89 & $<10^{-4}$  & 0.39 \\
$L_{\rm iso,52}$($E_{\rm p,z}$, $\Gamma_0$[Wind]) &  log$L_{\rm iso,52}$ = (-4.61 $\pm$ 0.30) + (0.70 $\pm$ 0.12) $\times$ log$E_{\rm p,z}$ + (1.67 $\pm$ 0.18) $\times$ log$\Gamma_0$ & 0.91 & $<10^{-4}$  & 0.35 \\
\hline
\textbf{Fitting results from all 89 GRBs:}\\
$\Gamma_0$[ISM]($E_{\rm iso,52}$) &  log$\Gamma_0$ = (2.00 $\pm$ 0.04) + (0.27 $\pm$ 0.03) $\times$ log$E_{\rm iso,52}$ & 0.74 & $<10^{-4}$  & 0.21 \\
$\Gamma_0$[Wind]($E_{\rm iso,52}$) &  log$\Gamma_0$ = (1.55 $\pm$ 0.02) + (0.35 $\pm$ 0.02) $\times$ log$E_{\rm iso,52}$ & 0.90 & $<10^{-4}$  & 0.14 \\
$\Gamma_0$[ISM]($L_{\rm iso,52}$) &  log$\Gamma_0$ = (2.17 $\pm$ 0.02) + (0.28 $\pm$ 0.02) $\times$ log$L_{\rm iso,52}$ & 0.80 & $<10^{-4}$  & 0.18 \\
$\Gamma_0$[Wind]($L_{\rm iso,52}$) &  log$\Gamma_0$ = (1.77 $\pm$ 0.02) + (0.33 $\pm$ 0.02) $\times$ log$L_{\rm iso,52}$ & 0.89 & $<10^{-4}$  & 0.15 \\
$\Gamma_0$[ISM]($E_{\rm p,z}$) &  log$\Gamma_0$ = (1.31 $\pm$ 0.15) + (0.36 $\pm$ 0.05) $\times$ log$E_{\rm p,z}$ & 0.59 & $<10^{-4}$  & 0.25 \\
$\Gamma_0$[Wind]($E_{\rm p,z}$) &  log$\Gamma_0$ = (0.69 $\pm$ 0.14) + (0.46 $\pm$ 0.05) $\times$ log$E_{\rm p,z}$ & 0.70 & $<10^{-4}$  & 0.23 \\
$E_{\rm iso,52}$($E_{\rm p,z}$, $\Gamma_0$[ISM]) &  log$E_{\rm iso,52}$ = (-3.94 $\pm$ 0.38) + (0.83 $\pm$ 0.12) $\times$ log$E_{\rm p,z}$ + (1.20 $\pm$ 0.20) $\times$ log$\Gamma_0$ & 0.84 & $<10^{-4}$  & 0.45 \\
$E_{\rm iso,52}$($E_{\rm p,z}$, $\Gamma_0$[Wind]) &  log$E_{\rm iso,52}$ = (-3.66 $\pm$ 0.22) + (0.41 $\pm$ 0.10) $\times$ log$E_{\rm p,z}$ + (1.88 $\pm$ 0.16) $\times$ log$\Gamma_0$ & 0.92 & $<10^{-4}$  & 0.33 \\
$L_{\rm iso,52}$($E_{\rm p,z}$, $\Gamma_0$[ISM])  &  log$L_{\rm iso,52}$ = (-5.21 $\pm$ 0.32) + (0.89 $\pm$ 0.10) $\times$ log$E_{\rm p,z}$ + (1.42 $\pm$ 0.17) $\times$ log$\Gamma_0$ & 0.90 & $<10^{-4}$  & 0.38 \\
$L_{\rm iso,52}$($E_{\rm p,z}$, $\Gamma_0$[Wind]) &  log$L_{\rm iso,52}$ = (-4.54 $\pm$ 0.23) + (0.62 $\pm$ 0.10) $\times$ log$E_{\rm p,z}$ + (1.73 $\pm$ 0.16) $\times$ log$\Gamma_0$ & 0.92 & $<10^{-4}$  & 0.33 \\
\enddata
\end{deluxetable*}
 
\clearpage
\renewcommand{\arraystretch}{1.2}
\begin{deluxetable*}{llccc}[h]
\centering
\tablecaption{The Results of the Regression Analysis for the Correlations in the GRBs Sample with Observed Jet Breaks, Where $r$ Is the Correlation Coefficient, $p$ Is the Chance Probability, and $\sigma_{\rm int}$ Is the Dispersion.
\label{tab:tab4}} 
\tablewidth{700pt}
\tabletypesize{\small}
\tablehead{
\multicolumn{1}{l}{Relations} & \multicolumn{1}{l}{Expressions} & \colhead{$r$} & \colhead{$p$} & \colhead{$\sigma_{\rm int}$} \\[-18pt]
}
\startdata
$E_{\rm iso}$($\theta_{\rm jet}$[ISM]) &  log$E_{\rm iso}$ = (51.88 $\pm$ 0.56) + (-0.92 $\pm$ 0.42) $\times$ log$\theta_{\rm jet}$ & -0.34 & 0.03 & 0.83 \\
$E_{\rm iso}$($\theta_{\rm jet}$[Wind]) &  log$E_{\rm iso}$ = (50.27 $\pm$ 0.44) + (-2.15 $\pm$ 0.33) $\times$ log$\theta_{\rm jet}$ & -0.73 & $<10^{-4}$ & 0.60 \\
$L_{\rm iso}$($\theta_{\rm jet}$[ISM]) &  log$L_{\rm iso}$ = (51.34 $\pm$ 0.61) + (-0.79 $\pm$ 0.45) $\times$ log$\theta_{\rm jet}$ & -0.28 & 0.08 & 0.90 \\
$L_{\rm iso}$($\theta_{\rm jet}$[Wind]) &  log$L_{\rm iso}$ = (49.75 $\pm$ 0.52) + (-2.00 $\pm$ 0.39) $\times$ log$\theta_{\rm jet}$ & -0.65 & $<10^{-4}$ & 0.71 \\
$E_{\rm p,z}$($\theta_{\rm jet}$[ISM]) & log$E_{\rm p,z}$ = (2.20 $\pm$ 0.34) + (-0.35 $\pm$ 0.25) $\times$ log$\theta_{\rm jet}$ & -0.23 & 0.15 & 0.50 \\
$E_{\rm p,z}$($\theta_{\rm jet}$[Wind]) &  log$E_{\rm p,z}$ = (1.41 $\pm$ 0.31) + (-0.95 $\pm$ 0.23) $\times$ log$\theta_{\rm jet}$ & -0.56 & 0.0001 & 0.42 \\
\hline
$\Gamma_0$[ISM]($\theta_{\rm jet}$[ISM]) &  log$\Gamma_0$ = (2.05 $\pm$ 0.20) + (-0.18 $\pm$ 0.15) $\times$ log$\theta_{\rm jet}$ & -0.19 & 0.22 & 0.30 \\
$\Gamma_0$[Wind]($\theta_{\rm jet}$[Wind]) &  log$\Gamma_0$ = (0.97 $\pm$ 0.19) + (-0.72 $\pm$ 0.14) $\times$ log$\theta_{\rm jet}$ & -0.65 & $<10^{-4}$ & 0.25 \\
\hline
$E_{\rm iso}$($\theta_{\rm jet}$[ISM], $\Gamma_0$[ISM])  &  log$E_{\rm iso}$ = (48.26 $\pm$ 0.68) + (-0.52 $\pm$ 0.26) $\times$ log$\theta_{\rm jet}$ + (2.21 $\pm$ 0.28) $\times$ log$\Gamma_0$ & 0.82 & $<10^{-4}$ & 0.51 \\
$E_{\rm iso}$($\theta_{\rm jet}$[Wind], $\Gamma_0$[Wind]) &  log$E_{\rm iso}$ = (49.39 $\pm$ 0.43) + (-0.65 $\pm$ 0.21) $\times$ log$\theta_{\rm jet}$ + (2.08 $\pm$ 0.19) $\times$ log$\Gamma_0$ & 0.95 & $<10^{-4}$ & 0.29 \\
$L_{\rm iso}$($\theta_{\rm jet}$[ISM], $\Gamma_0$[ISM])  &  log$L_{\rm iso}$ = (46.90 $\pm$ 0.70) + (-0.35 $\pm$ 0.27) $\times$ log$\theta_{\rm jet}$ + (2.46 $\pm$ 0.29) $\times$ log$\Gamma_0$ & 0.83 & $<10^{-4}$ & 0.52 \\
$L_{\rm iso}$($\theta_{\rm jet}$[Wind], $\Gamma_0$[Wind]) &  log$L_{\rm iso}$ = (47.95 $\pm$ 0.57) + (-0.29 $\pm$ 0.28) $\times$ log$\theta_{\rm jet}$ + (2.37 $\pm$ 0.25) $\times$ log$\Gamma_0$ & 0.91 & $<10^{-4}$ & 0.39 \\
$E_{\rm p,z}$($\theta_{\rm jet}$[ISM], $\Gamma_0$[ISM])  &  log$E_{\rm p,z}$ = (0.19 $\pm$ 0.51) + (-0.15 $\pm$ 0.19) $\times$ log$\theta_{\rm jet}$ + (1.11 $\pm$ 0.21) $\times$ log$\Gamma_0$ & 0.69 & $<10^{-4}$ & 0.38 \\
$E_{\rm p,z}$($\theta_{\rm jet}$[Wind], $\Gamma_0$[Wind]) &  log$E_{\rm p,z}$ = (0.64 $\pm$ 0.48) + (-0.17 $\pm$ 0.23) $\times$ log$\theta_{\rm jet}$ + (1.09 $\pm$ 0.21) $\times$ log$\Gamma_0$ & 0.78 & $<10^{-4}$ & 0.32 \\
\enddata
\end{deluxetable*}

\clearpage
\renewcommand{\arraystretch}{1.2}
\begin{deluxetable*}{llccc}[h]
\centering
\tablecaption{The Results of the Regression Analysis for the Correlations of $E_{\gamma}$, $L_{\gamma}$, and $E_{\rm p,z}$ with $\Gamma_{0}$, Where $r$ Is the Correlation Coefficient, $p$ Is the Chance Probability, and $\sigma_{\rm int}$ Is the Intrinsic Dispersion.
\label{tab:tab5}} 
\tablewidth{700pt}
\tabletypesize{\small}
\tablehead{
\multicolumn{1}{l}{Relations} & \multicolumn{1}{l}{Expressions} & \colhead{$r$} & \colhead{$p$} & \colhead{$\sigma_{\rm int}$} \\[-18pt]
}
\startdata
$\Gamma_0$[ISM]($E_{\gamma,52}$[ISM]) &  log$\Gamma_0$ = (2.69 $\pm$ 0.09) + (0.22 $\pm$ 0.04) $\times$ log$E_{\gamma,52}$ & 0.64 & $<10^{-4}$ & 0.23 \\
$\Gamma_0$[Wind]($E_{\gamma,52}$[Wind]) &  log$\Gamma_0$ = (2.65 $\pm$ 0.12) + (0.40 $\pm$ 0.06) $\times$ log$E_{\gamma,52}$ & 0.72 & $<10^{-4}$ & 0.23 \\
$\Gamma_0$[ISM]($L_{\gamma,52}$[ISM]) &  log$\Gamma_0$ = (2.81 $\pm$ 0.10) + (0.20 $\pm$ 0.04) $\times$ log$L_{\gamma,52}$ & 0.66 & $<10^{-4}$ & 0.23 \\
$\Gamma_0$[Wind]($L_{\gamma,52}$[Wind]) &  log$\Gamma_0$ = (2.69 $\pm$ 0.16) + (0.30 $\pm$ 0.06) $\times$ log$L_{\gamma,52}$ & 0.64 & $<10^{-4}$ & 0.26 \\
\hline
$E_{\gamma,52}$[ISM]($E_{\rm p,z}$, $\Gamma_0$[ISM])  &  log$E_{\gamma,52}$ = (-6.22 $\pm$ 0.82) + (0.65 $\pm$ 0.29) $\times$ log$E_{\rm p,z}$ + (1.16 $\pm$ 0.48) $\times$ log$\Gamma_0$ & 0.70 & $<10^{-4}$  &  0.65 \\
$E_{\gamma,52}$[Wind]($E_{\rm p,z}$, $\Gamma_0$[Wind])  &  log$E_{\gamma,52}$ = (-4.41 $\pm$ 0.41) + (0.24 $\pm$ 0.21) $\times$ log$E_{\rm p,z}$ + (1.01 $\pm$ 0.33) $\times$ log$\Gamma_0$ & 0.73 & $<10^{-4}$  & 0.42 \\
$L_{\gamma,52}$[ISM]($E_{\rm p,z}$, $\Gamma_0$[ISM])  &  log$L_{\gamma,52}$ = (-7.43 $\pm$ 0.87) + (0.71 $\pm$ 0.31) $\times$ log$E_{\rm p,z}$ + (1.31 $\pm$ 0.52) $\times$ log$\Gamma_0$ & 0.71 & $<10^{-4}$  & 0.70 \\
$L_{\gamma,52}$[Wind]($E_{\rm p,z}$, $\Gamma_0$[Wind])  &  log$L_{\gamma,52}$ = (-5.34 $\pm$ 0.52) + (0.43 $\pm$ 0.27) $\times$ log$E_{\rm p,z}$ + (0.86 $\pm$ 0.42) $\times$ log$\Gamma_0$ & 0.67 & $<10^{-4}$  & 0.54 \\
\enddata
\end{deluxetable*}

\end{document}